\documentclass[]{mnras}
\usepackage[utf8x]{inputenc}
\usepackage{float}
\usepackage{graphicx}
\usepackage{amssymb}
\usepackage{amsmath}
\usepackage[usenames,dvipsnames]{xcolor}
\usepackage{multicol}
\usepackage{multirow}
\usepackage{enumerate}
\usepackage{natbib}
\usepackage{mathtools}

\DeclarePairedDelimiter\abs{\lvert}{\rvert}%
\DeclarePairedDelimiter\norm{\lVert}{\rVert}%

\makeatletter
\let\oldabs\abs
\def\abs{\@ifstar{\oldabs}{\oldabs*}}
\let\oldnorm\norm
\def\norm{\@ifstar{\oldnorm}{\oldnorm*}}
\makeatother

\newcommand{\rvir}{R_\mathrm{v}}

\newcommand{\vmax}{V_\mathrm{max}}

\newcommand{\msun}{M_\odot}
\newcommand{\mstar}{M_\star}
\newcommand{\mhalo}{M_\mathrm{halo}}
\newcommand{\mpeak}{M_\mathrm{peak}}

\newcommand{\kpc}{\mathrm{kpc}}

\newcommand{\kms}{{\rm km} \, {\rm s}^{-1}}

\newcommand{\vcirc}{V_\mathrm{circ}}
\newcommand{\rmax}{R_\mathrm{max}}
\newcommand{\vhalf}{V_\mathrm{1/2}}
\newcommand{\rhalf}{R_\mathrm{1/2}}

\defcitealias{Behroozi2013}{B13}

\title[Organized Chaos:  Scatter in $\mstar$ -- $\mhalo$]{Organized Chaos: Scatter in the relation between stellar mass and halo mass in small galaxies}
\author[S. Garrison-Kimmel et al.]{Shea Garrison-Kimmel$^{1}$\thanks{$\!$sheagk@caltech.edu},
  James S. Bullock$^{2}$,
  Michael Boylan-Kolchin$^{3}$, \and
  Emma Bardwell$^{4}$ \\
  \noindent$\!\!$ $^1$TAPIR, California Institute of Technology, Pasadena, CA 91125, USA \\
    \noindent$\!\!$ $^2$Center for Cosmology, Department of Physics and Astronomy,
  University of California, Irvine, CA 92697, USA \\
  \noindent$\!\!$ $^3$Department of Astronomy, The University of Texas at Austin, 2515 Speedway, Stop C1400, Austin, TX 78712 \\
  \noindent$\!\!$ $^4$Case Western Reserve University, Cleveland, Ohio 44106, USA \\
}
\begin{document}

 \pagerange{\pageref{firstpage}--\pageref{lastpage}}
 \pubyear{2014}

\maketitle

\label{firstpage}
\begin{abstract}
  We use Local Group galaxy counts together with the ELVIS N-body simulations to
  explore the relationship between the scatter and slope in the stellar mass
  vs. halo mass relation at low masses, $\mstar \simeq 10^5 - 10^8 \msun$.
  Assuming models with log-normal scatter about a median relation of the form
  $\mstar \propto \mhalo^\alpha$, the preferred log-slope steepens from
  $\alpha \simeq 1.8$ in the limit of zero scatter to $\alpha \simeq 2.6$ in the
  case of $2$~dex of scatter in $\mstar$ at fixed halo mass.  We provide fitting
  functions for the best-fit relations as a function of scatter, including cases
  where the relation becomes increasingly stochastic with decreasing
  mass.  We show that if the scatter at fixed halo mass is large enough
  ($\gtrsim 1$~dex) {\em and} if the median relation is steep enough
  ($\alpha \gtrsim 2$), then the ``too-big-to-fail'' problem seen in the Local
  Group can be self-consistently eliminated in about $\sim 5-10\%$ of
  realizations.  This scenario requires that the most massive subhalos host
  unobservable ultra-faint dwarfs fairly often; we discuss potentially
  observable signatures of these systems.  Finally, we compare our derived
  constraints to recent high-resolution simulations of dwarf galaxy formation in
  the literature.  Though simulation-to-simulation scatter in $\mstar$ at fixed
  $\mhalo$ is large among separate authors ($\sim 2$~dex), individual codes produce
  relations with much less scatter and usually give relations that would over-produce
  local galaxy counts.
 \end{abstract}


\section{Introduction}
\label{sec:intro}

The standard paradigm of galaxy formation posits that dark matter halos are the sites of
galaxy formation \citep{White78,Blumenthal84}.
The underlying relationship between galaxies and dark matter halos has emerged
as an important benchmark for detailed theories.  In particular, reproducing the
observed abundance of galaxies as a function of stellar mass or luminosity
imposes a particularly powerful constraint on models
\citep{Klypin1999,Benson2000,Bullock2002,Berlind2002,Kravtsov2004}.

Perhaps the simplest example of this approach is to assume that more
massive dark matter halos host brighter galaxies, such that the cumulative
number density of dark matter halos matches the cumulative number density of
galaxies in a nearly one-to-one fashion.  On large scales and for bright galaxies,
this abundance matching (AM) assumption, when applied using cosmological,
DM-only simulations, yields clustering statistics that agree remarkably
well with observations for $\sim L_*$ galaxies with $\mstar \simeq 10^{10-11}$
\citep{Conroy2006,Reddick2013}.  Moreover, observational estimates for halo masses,
e.g. from X-ray temperatures or virial-scale satellite kinematics, agree well with AM
relations up to the scale of giant cluster galaxies $\mstar \simeq 10^{12} \msun$
\citep[e.g.][]{Kravtsov2014}.  AM even appears to work well down to
$\mstar\sim10^9\msun$:  when applied to the Millennium-II simulation \citep{MSII},
it correctly reproduces not only the fraction of Milky Way (MW)-size hosts with
LMC-size satellites, but also the radial and velocity distributions of those
satellites \citep{Tollerud2011}.  Perhaps most surprisingly,  galaxies over this
range ($\mstar \simeq 10^{9-12} \msun $) appear to populate halos with a fairly
small amount of scatter at fixed halo mass.  \citet{Behroozi2013}, for example,
report a log-normal scatter (one standard deviation) of $0.2$~dex in $\mstar$ at
fixed $\mhalo$.  \citet{Reddick2013} also prefer $0.2$~dex of scatter based on
their clustering analysis.  \citet{Kravtsov2014} quote $0.1$~dex for brightest
cluster galaxies.

The focus of the present work is to explore the relationship between halos
and galaxies below the stellar mass range that is easily accessible in large,
volume-limited surveys: $\mstar < 10^8 \msun$.   Simply extrapolating the slope
of the relationship reported at higher masses would suggest that dwarf galaxies
obey a relation $\mstar \propto \mhalo^\alpha$, with
$\alpha \simeq 1.6-1.9$ \citep[][]{Behroozi2013,Moster2013}.  In fact, \citet{ELVIS}
found that this assumption with $\alpha = 1.9$ and minimal scatter does reproduce the
abundance of Local Group (LG) galaxies down to  $\mstar \simeq 10^{6} \msun$ reasonably
well.  In contrast, \citet{Brook2014} performed a similar analysis and found a preferred
log-slope of $\alpha = 3.1$ over the tight mass range $\mstar \simeq 10^{6-8} \msun$. This
is much steeper than the relationship one expects from extrapolating the trend seen at higher
masses.  However, the log-slope obtained by \citet{Brook2014} is degenerate with the
normalization at high masses ($\mstar \sim 10^{9.5}\msun$), which, due to their method
of matching simulated halo mass functions to the observed stellar mass function, is sensitive
to the masses of the largest halos that form in their Local Group environments.  In fact,
\citet{Brook2014} predict very similar halo masses to the \citet{ELVIS} relation for
$\mstar \sim 10^6 - 10^7 \msun$.

While AM (with minimal scatter) can reasonably reproduce the counts of local
galaxies, there are reasons to consider the possibility that it breaks down
entirely at very low masses, or at least that the scatter increases
dramatically.  One motivation comes from galaxy formation physics, which
suggests that star formation and gas retention in small halos may be very
sensitive to small changes in the halo's virial temperature and mass-accretion
history reaching back to the epoch of reionization
\citep{Bullock2000,Somerville2002,Onorbe2015,Sawala2014}. Another clue comes from
direct mass measurements in small galaxies themselves.  The smallest satellite galaxies
of the MW demonstrate no observable trend between central dynamical mass
and luminosity \citep{Strigari2008}, which may suggest that there is no trend
between total halo mass and stellar mass (or at least that the relationship is
very steep).  A related result was presented by \citet{TBTF1,TBTF2}, who
compared the central masses of dark mater subhalos in the ultra-high resolution
Aquarius simulations of MW-size hosts \citep{Aquarius} to the masses
of the bright MW dwarf spheroidals (dSphs).  They found that the largest
subhalos are denser than any of the known MW dwarfs.  One interpretation could
be that the most massive predicted subhalos have simply failed to form enough
stars for them to be observable, but given the comparatively high mass of the
offending halos (well above the mass where photoionization is expected to act)
we expect the missing halos to be ``too big to fail" (TBTF).

A wide variety of solutions for TBTF have been proposed.  For example,
\citet{diCintio2011}, \citet{Wang2012}, \citet{TBTF2}, and \citet{Vera-Ciro2013}
pointed out that reducing the assumed mass of the MW host halo to
$\mhalo\sim8\times10^{11}\msun$ will reduce the number of problematic subhalos.
It is also possible that the substructure population of the MW is abnormal for
its mass \citep{Purcell2012,Rodriguez2013,ELVISTBTF}, though the fact that the
same problem appears to exist around M31 \citep{Tollerud2014} makes such a
solution less appealing.  The detailed analysis of \citet{Jiang2015} strongly
suggests that such a case would be exceedingly rare.

Others \citep[e.g.][]{Zolotov2012,Brooks2013,Arraki2012,DelPopolo2014}
have argued that interactions with the baryons in the host galaxy, which
are not included in dark-matter only simulations, act to reduce the
central densities of the massive failures such that their integrated
masses within $\sim1$~kpc are similar to those of the MW dSphs.
Similarly, \citet{Read2006} used idealized N-body simulations to
demonstrate that tidal shocking may lower the central velocity
dispersions of dwarf halos if the halos are accreted with cored
profiles.  These solutions, however, are only applicable within
the virial radius ($\rvir$) of the MW, and break down when faced with the
ubiquity of TBTF not only in the nearby field \citep{ELVISTBTF},
but also well beyond the LG \citep{Papastergis2015,Papastergis2016}.

Other authors have proposed explanations for TBTF that do not rely
on environmental influences, and which could therefore act even
in the field.  Gas blowouts, driven by supernovae events within the dwarfs,
may add energy to dark matter particle orbits and reduce the central
densities enough to match observations
\citep{Navarro1996,Mashchenko2008,Pontzen2012,Governato2012,BrooksZolotov2012,Amorisco2013feedback,Gritschneder2013}.
Though many authors \citep{Gnedin2002,Penarrubia2012,GK2013,diCintio2014} have
argued that the galaxies of interest ($\mstar\lesssim10^6\msun$) do
not produce enough supernovae to drive the required blowouts, other authors
\citep[e.g.][]{Read2005,Maxwell2015} have made the point that energetic arguments
depend crucially on just how the dark matter particle orbits are perturbed.
\citet{Onorbe2015} presented a single simulation of a $\mstar\sim2\times10^6\msun$
galaxy that formed within a TBTF halo with a central core density consistent with
those observed for similarly sized dwarfs.  \citet{Read2016a,Read2016b} used
idealized simulations to argue that these cores form in halos as small as
$\mhalo = 10^8\msun$, though their simulations form $\sim10$ times more stars
than AM extrapolations predict.  It is not clear whether core formation of
this kind should be common enough to resolve the problem generally.

More recently, \citet{Papastergis2016} have used HI data to show that the mass deficit
exists at radii well beyond the stellar extent of the galaxy.  This is particularly hard
to reconcile in models where stellar-feedback driven core formation solves TBTF in
energetically feasible ways \citep{diCintio2014,Onorbe2015,Maxwell2015,Chan2015}.
This is because the dark matter cores produced by supernovae explosions in these cosmological
simulations are typically similar in size to the stellar half-light radius of the
galaxy.~\footnote{\citet{GK2013} used simple numerical experiments to explore models that
removed dark matter mass well beyond the galaxy half-light radius, but showed that these
cases require even more energy, likely exceeding the energy budget available from supernovae.}

Concerns about the ability of baryonic physics to resolve TBTF have motivated
solutions that are cosmological in nature. These include modifications to the
small-scale power spectrum \citep{Polisensky2013,GK2014} and altering the
assumption that dark matter is both cold
\citep[][]{Anderhalden2013,Lovell2013,Horiuchi2016,Bozek2016} and
collisionless \citep{Vogelsberger2012,Rocha2013,Zavala2013,Elbert2014}.

Most of the explanations detailed above alter the density profiles
of massive (sub)halos such that they may host the observed dwarfs.
In principle, however, these massive dwarf halos may remain cuspy
and dense but form far fewer stars than simple, low-scatter extrapolations
of AM relations suggest.  In such a situation, a TBTF halo would not be
``too big to fail" after all, but actually remain undetectable to current
observations.  In practice, therefore, TBTF may be indicative of increased
scatter in the $\mstar$-$\mhalo$ relation at small masses.  If so, this would
indicate that galaxy formation proceeds in a fundamentally distinct way
in dark matter halos smaller than $\mhalo\sim10^{11}\msun$, perhaps because
the baryons in those potential wells are far more susceptible to both internal
and external perturbations than previously expected.  Understanding the scatter
in the $\mstar$-$\mhalo$ relation is thus crucial to our overall understanding
of how stars and galaxies form in dark matter halos.

In what follows, we use the ELVIS (Exploring the Local Volume in Simulations)
suite \citep{ELVIS} along with the stellar mass function (SMF) of galaxies
in the Local Group (LG) to constrain the $\mstar$-$\mhalo$ relation down to the
lowest stellar masses possible due to completeness $\mstar \simeq 10^5 \msun$.  We
normalize our relation at high masses to the results of \citet[hereafter B13]{Behroozi2013}
and explore models with a variable faint-end log-slope $\alpha$ and a log-normal scatter
$\sigma$ and explore the covariance between them.

This paper is organized as follows:  \S\ref{sec:methods} discusses
the observational and theoretical data employed here, followed
by the methods for assigning stellar masses to (sub)halos and
quantifying the goodness of fit to the $\mstar$ functions.
\S\ref{sec:results} presents the best-fit log-slope, and explores
one impact of varying the scatter by investigating the severity of TBTF,
quantified by the number of ``bright'' massive failures in each
realization.  We  close by discussing other observable consequences
of an increase in the low mass scatter in \S\ref{sec:discussion} and
briefly summarize in \S\ref{sec:conclusions}.

\section{Methods}
\label{sec:methods}

\subsection{Observational Data}
\label{ssec:obs}

In order to maximize the observational sample size, we adopt three independent
completeness cuts for dwarfs that are either 1) satellites of the Milky Way;
2) satellites of M31; or 3) non-satellite that exist within in the Local Group.
We include all eleven classical satellites of the MW, defined as those identified
prior to SDSS within $300$~kpc of the Galactic center.  The faintest of these, Draco,
defines our incompleteness limit of $\mstar = 4.5\times10^5\msun$.
Completeness around M31 is perhaps better understood due to the
uniformity of the PAndAS survey~--~Figure~24 of \citet{Tollerud2012}
illustrates that PAndAS is complete to galaxies with half-light luminosities
$L_{1/2} > 10^5L_\odot$, for typical half-light radii.\footnote{If, however, more
spatially extended galaxies ($\rhalf > 1~\kpc$) are common, then we are under-estimating
the completeness, and would therefore over-estimate the steepness of the $\mstar-\mhalo$
relation.}
Assuming a mass-to-light ratio of two,
consistent with the findings of \citet{Martin2008} and with extrapolations of
the \citet{Bell2001} relations to the dwarf scale, and including the total
stellar mass of the galaxy, we therefore adopt an incompleteness limit of
$\mstar = 10^5\msun$, but only include galaxies (and subhalos) within the PAndAS footprint,
the central $150$~kpc of the galaxy.  In the ``Local Field'' (LF), defined here as the region
more than 300 kpc from both the MW and M31, but within 1.2~Mpc of either (a cut chosen to
match the possible cuts in the simulations; see \S\ref{ssec:sims}),  we adopt
an incompleteness limit of $\mstar = M_\mathrm{\star,\,Cetus} = 4.5\times10^6\msun$.
We acknowledge, however, that incompleteness in the LF is poorly understood, and we
therefore analyze the LF separately from the satellite populations.

We exclude the two most massive satellites of both the MW (the LMC and the SMC)
and M31 (M33 and NGC~205) from our analysis because such luminous satellites are
known to be rare around MW-size systems \citep{Boylan-Kolchin2010,Busha2011,Tollerud2011},
and many of the ELVIS hosts lack the massive subhalos that appear to host these
systems \citep{Tollerud2011};  including these systems, however, has only a slight
effect on the results.  Halos assigned stellar masses below the completeness limit
are not considered in calculating the goodness-of-fit between the observational
SMFs and those derived from the simulations (see \S\ref{ssec:kappa}).  Many models
therefore overproduce counts below the cited completeness limits.  If there are
undiscovered galaxies \emph{above} the completeness limits, then we will underestimate
the best-fit log-slope.

\begin{figure}
\includegraphics[width=\columnwidth]{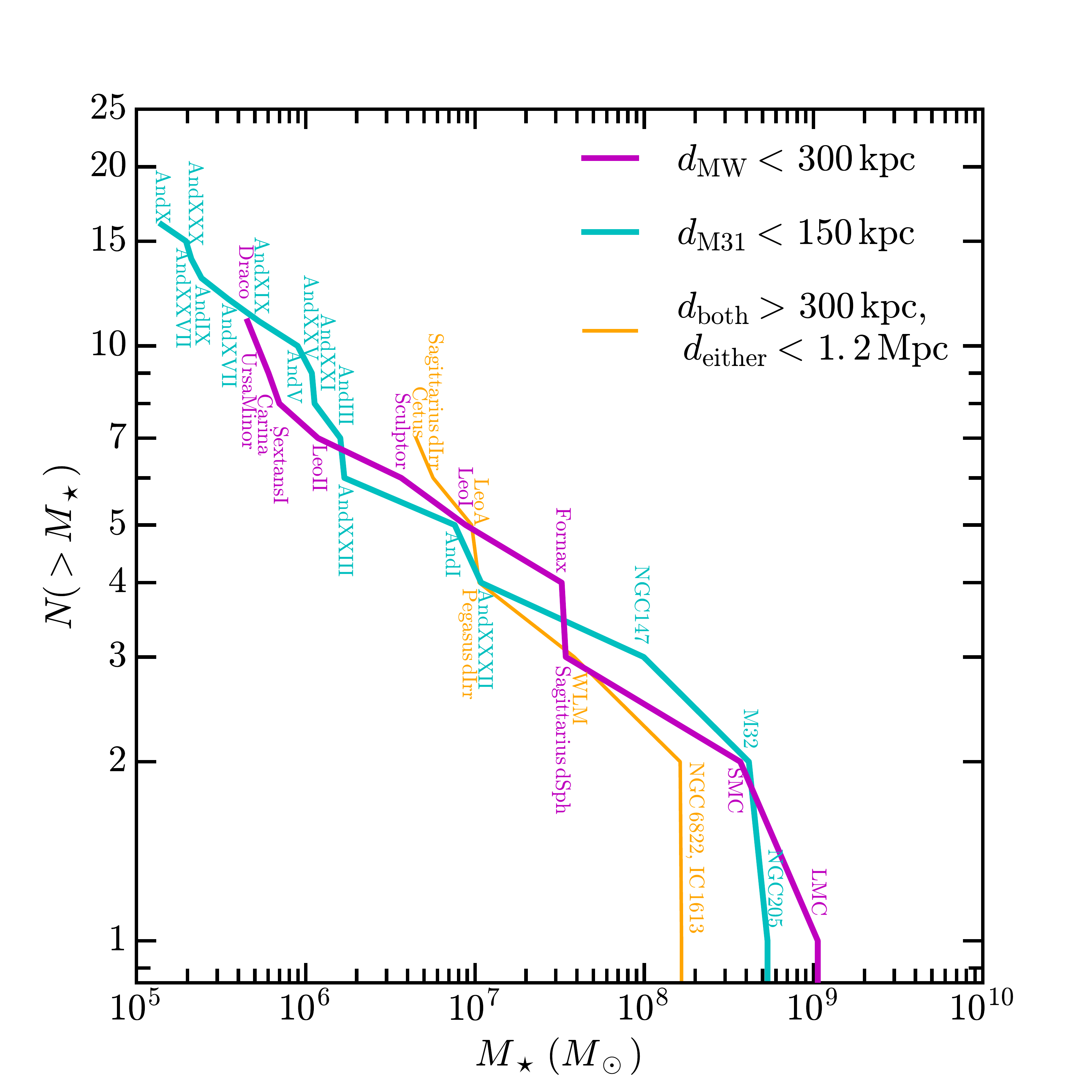}
\caption{The stellar mass functions of the classical MW satellites
(magenta), the M31 satellites within the PAndAS survey volume
($d_\mathrm{M31} < 150$~kpc, cyan), and galaxies in the Local
Field (orange), which includes all galaxies more than $300$~kpc from
both the MW and M31, but within 1.2~Mpc of either giant.  Stellar
masses are from \citet{Woo2008}, where available, and are otherwise
calculated from the luminosities tabulated in \citet{McConnachie2012},
assuming a mass-to-light ratio of two.  Each histogram is truncated at
the adopted completeness limit:
$\mstar = M_\mathrm{\star,\,Draco} = 4.5\times10^{5}\msun$ for the
MW satellites, $\mstar = 10^5\msun$ for satellites around M31
\citep{Tollerud2012}, and $\mstar = 4.5\times10^6\msun$ in the Local Field.}
\label{fig:obsmstar}
\end{figure}

The names, stellar masses, and adopted distances from the centers of the MW
and M31 are listed in Table~\ref{tab:obsdata}; we explicitly note that our
distance cuts exclude M33 (Triangulum), which lies outside the PAndAS footprint.
The anti-cumulative stellar mass functions are also plotted in
Figure~\ref{fig:obsmstar} for the MW (magenta), M31 (cyan), and in the LF (orange),
with the individual galaxies labeled.  Stellar masses are drawn from
\citet{Woo2008} where possible, and are otherwise calculated from the data in
\citet{McConnachie2012}, assuming $\mstar/L = 2$ .  Similarly, the
MW-centric distances are taken from \citet{McConnachie2012}, and the
M31-centric distances are calculated from the positions given in that
work, assuming an MW-M31 separation of $787$~kpc \citep{McConnachie2005}.

\begin{table}
\centering
\begin{tabular}{lccc}
\hline
Name & $M_{\star}$ & $d_{\rm MW}$ & $d_{\rm M31}$ \\
& ($\msun$) & (kpc) & (kpc) \\
\hline\hline
\multicolumn{4}{c}{\textbf{MW satellites ($\mathbf{d_\mathrm{MW} < 300}$~kpc)}} \\
LMC & 1.06$\times10^9$ & 50 & 811 \\
SMC & 3.69$\times10^8$ & 61 & 812 \\
Sagittarius dSph & 3.44$\times10^7$ & 19 & 792 \\
Fornax & 3.25$\times10^7$ & 149 & 773 \\
Leo I & 8.79$\times10^6$ & 257 & 922 \\
Sculptor & 3.67$\times10^6$ & 86 & 766 \\
Leo II & 1.18$\times10^6$ & 236 & 902 \\
Sextans I & 6.98$\times10^5$ & 89 & 839 \\
Carina & 6.03$\times10^5$ & 107 & 842 \\
Ursa Minor & 4.53$\times10^5$ & 78 & 758 \\
Draco & 4.53$\times10^5$ & 76 & 755 \\
\multicolumn{4}{c}{\textbf{M31 satellites ($\mathbf{d_\mathrm{M31} < 150}$~kpc)}} \\
NGC 205 & 5.35$\times10^8$ & 828 & 42 \\
M32 & 4.15$\times10^8$ & 809 & 23 \\
NGC 147 & 9.91$\times10^7$ & 680 & 143 \\
Andromeda XXXII & 1.09$\times10^7$ & 780 & 141 \\
Andromeda I & 7.59$\times10^6$ & 749 & 58 \\
Andromeda XXIII & 1.69$\times10^6$ & 774 & 126 \\
Andromeda III & 1.6$\times10^6$ & 752 & 75 \\
Andromeda XXI & 1.13$\times10^6$ & 831 & 134 \\
Andromeda XXV & 1.09$\times10^6$ & 817 & 89 \\
Andromeda V & 8.96$\times10^5$ & 777 & 110 \\
Andromeda XIX & 5.3$\times10^5$ & 824 & 114 \\
Andromeda XVII & 3.47$\times10^5$ & 732 & 70 \\
Andromeda IX & 2.42$\times10^5$ & 770 & 41 \\
Andromeda XXX & 2.11$\times10^5$ & 686 & 148 \\
Andromeda XXVII & 1.96$\times10^5$ & 832 & 74 \\
Andromeda X & 1.41$\times10^5$ & 674 & 139 \\
\multicolumn{4}{c}{\textbf{Local Field Galaxies}} \\
NGC 6822 & 1.66$\times10^8$ & 452 & 898 \\
IC 1613 & 1.63$\times10^8$ & 758 & 520 \\
WLM & 3.86$\times10^7$ & 933 & 836 \\
Pegasus dIrr & 1.06$\times10^7$ & 921 & 474 \\
Leo A & 9.55$\times10^6$ & 803 & 1200 \\
Sagittarius dIrr & 5.65$\times10^6$ & 1059 & 1357 \\
Cetus & 4.49$\times10^6$ & 756 & 680 \\
\end{tabular}
\caption{The names of galaxies included in this work, along with their
adopted stellar masses and distances from the MW and M31.  The sample
includes the classical satellites of the MW, dwarfs with $\mstar\geq10^5\msun$
within $150$~kpc of M31, and galaxies brighter than Cetus in the
LF.  Stellar masses are taken from \citet{Woo2008} where available, and
are otherwise drawn from the data compiled in \citet{McConnachie2012},
assuming a mass-to-light ratio of two.  MW-centric distances are also drawn
from \citet{McConnachie2012} and M31-centric distances are calculated
assuming a MW-M31 separation of $787$~kpc \citep{McConnachie2005}.}
\label{tab:obsdata}
\end{table}

\subsection{Simulations}
\label{ssec:sims}

Subhalo and field halo mass functions are drawn from the Exploring the Local
Volume In Simulations (ELVIS) suite, detailed in full in \citet{ELVIS}.  Briefly,
ELVIS includes twelve LG-like MW/M31 pairs of hosts, selected to be $\sim800$~kpc
apart, moving towards one another, and isolated from any larger systems.  ELVIS
additionally includes twenty-four isolated hosts, mass-matched to the systems
in the pairs.  All systems are free of contaminating low-resolution (high mass)
particles within their virial radii.  Moreover, the fields around the paired
halos are uncontaminated within $\gtrsim1$~Mpc of each halo center:  only four
pairs contain low resolution particles within $1.2$~Mpc of a host center, and
the low-resolution mass fraction within that volume is $\ll0.1\%$ in each of
those systems.  We therefore define the ``Local Field'' (LF) as the volume within
$1.2$~Mpc of either host center, but more than $300$~kpc from both hosts
\citep[as in][]{ELVIS}.

The ELVIS suite was initialized using \texttt{MUSIC} \citep{MUSIC} and simulated
using a combination of \texttt{GADGET-2} and \texttt{GADGET-3} \citep{GADGET} assuming
WMAP-7 cosmological parameters \citep{Larson2011}:  $\Omega_m = 0.266$, $\Omega_\Lambda = 0.734$,
$n_s = 0.963$, $\sigma_8 = 0.801$, and $h = 0.71$.  All forty-eight halos
were simulated with high-resolution regions embedded within boxes $70.4$~Mpc
on a side.  The particle mass $m_p$ within that high-resolution volume is
$1.9\times10^5\msun$; the Plummer-equivalent force softening $\epsilon$ is
held fixed in comoving units for $z > 9$, after which $\epsilon$ becomes
$141$~pc in physical units.\footnote{Three of the isolated halos were also
simulated with eight times more particles in the high-resolution regions
($m_p = 2.35\times10^4\msun$, $\epsilon = 70.4$~pc).  \citet{ELVIS} used
these simulations to show that the ELVIS suite is complete to
$\mpeak = 6\times10^7\msun$, which is well below the adopted observational
completeness cut for the vast majority of the models that we explore here:
for $\alpha = 2$, e.g., $\mstar(\mpeak=6\times10^7\msun) \sim 10^2~\msun$.
However, numerical incompleteness in the halo catalog may influence our results
for large scatters and/or very shallow slopes.  We therefore calculate the
best-fit log-slopes, derived from comparing realizations drawn from those
three isolated halos to the stellar mass function of the MW, both at the
fiducial resolution and at higher resolution.  Averaged over $500$
realizations, the best-fit $\alpha$ differs by $\sim6\%$ at most,
indicating that our results are independent of numerical resolution.}

Following \citetalias{Behroozi2013}, we parameterize the $\mstar$-$\mhalo$
relation in terms of $\mpeak$, the largest instantaneous virial mass
associated with the main branch of each (sub)halo's merger tree.
Particularly for subhalos, which have been significantly stripped via
interactions with the MW/M31 host, ``peak'' quantities appear to correlate
more strongly with $\mstar$ than the subhalo properties today
\citep{Reddick2013}.  Peak quantities for the ELVIS halos are drawn from
merger trees created with \texttt{rockstar} \citep{rockstar} and
\texttt{consistent-trees} (\citealt{ctrees}; for more details, see \citealt{ELVIS}).

In order to create an analogous comparison set to the observational data,
we create satellite populations within both $150$~kpc and $300$~kpc of each
ELVIS halo center for every combination of $\alpha$ and $\sigma$.  The
former (latter) population is then compared to the M31 (MW) dataset.  To
find a final goodness-of-fit, each ELVIS pair is compared to the paired
observational dataset, where the better fit matching between the two halos
and the MW/M31 is used in each realization.  Because the isolated ELVIS
hosts have mass functions that are indistinguishable from those of the paired hosts
\citep{ELVIS}, we create ``pairs'' from the isolated hosts to maximize our simulation
sample; the isolated halos, however, are excluded from the LF analysis.  Furthermore,
the fields around three of the ELVIS pairs (Siegfried \& Roy,
Serena \& Venus, and Kek \& Kauket) are systematically offset to
much higher masses at fixed number, compared to the remaining nine pairs.
These systems thus prefer extremely steep log-slopes ($\alpha \gtrsim 4$) at
all scatters to avoid overproducing the LF SMF.  We therefore eliminate them
from our LF analysis because they do not appear to be representative
of the field around the MW.

\subsection{Assigning stellar masses to halos}
\label{ssec:models}

We explore monotonic relationships between $\mstar$ and $\mpeak$
with log-normal symmetric scatter $\sigma$ in $\mstar$ at fixed $\mpeak$.
The models match the results of \citetalias{Behroozi2013} above the
dwarf scale:  for $\mpeak > M_1 \sim 10^{11.5}\msun$ (roughly the mass
at which the low-mass slope begins to dominate), we directly adopt
the median relation in \citetalias{Behroozi2013}.  We emphasize that
the stellar mass predicted at $M_1$, $\mstar\sim 5\times10^9\msun$,
roughly matches the stellar mass of the LMC, $\sim10^9\msun$.
As noted above, the population statistics of LMC-size galaxies
are reproduced well with minimal scatter in AM relations, but
both simulations and observations suggest that different physical
processes may become important on smaller scales.  We therefore allow
both the log-slope and the scatter to vary freely for $\mpeak < M_1$.

Specifically, \citetalias{Behroozi2013} parameterizes $\mstar(\mhalo)$
as
\begin{equation}
\log_{10} \mstar = \log_{10}(\epsilon M_1) + f(\log_{10}(\frac{\mhalo}{M_1})) - f(0)
\label{eqn:behroozi1}
\end{equation}
where
\begin{equation}
f(x) = -\log_{10}(10^{-\alpha x}+1)+\delta \frac{(\log_{10}(1+\exp(x)))^\gamma}{1+\exp(10^{-x})}.
\label{eqn:behroozi2}
\end{equation}
A full interpretation of the parameters, which vary with redshift, is given in
Table 1 of \citetalias{Behroozi2013}; briefly, $M_1$ and $\epsilon$ indicate the
characteristic halo mass and stellar mass to halo mass ratio, respectively, while
$\gamma$ and $\delta$ govern the behavior at the high mass end ($\mhalo > M_1$).
For our purposes, however, the low mass end is most important, where
Equation~\ref{eqn:behroozi1} behaves as a power-law with log-slope $\alpha$:
$\mstar\propto\mhalo^\alpha$.  In the analysis that follows, we hold the parameters
quoted in Section 5 of \citetalias{Behroozi2013} fixed for $\mhalo > M_1$, but replace
$\alpha$ with values ranging freely from $0$ to $4$ for $\mhalo < M_1$.
For comparison, \citetalias{Behroozi2013} derive a faint-end log-slope
$\alpha = 1.412$ at $z = 0$, though \citet{ELVIS} showed that $\alpha = 1.92$
better reproduces galaxy counts in the LG with the same normalization,
assuming zero scatter.  Below, however, we demonstrate explicitly that
the best-fit $\alpha$ varies monotonically with the assumed scatter.

We primarily explore two models for including scatter in the
$\mstar-\mhalo$ relation.  The first, which we refer to as
``constant scatter,'' assumes a symmetric, log-normal scatter
$\sigma$ in $\mstar$ at fixed $\mpeak$ that is independent of
$\mpeak$.  This model is based on the results of
\citetalias{Behroozi2013}, who quote a constant $\sigma$ of $0.2$~dex
for $\mstar > 10^{8.5}\msun$.  Rather than introducing a
discontinuity in the scatter at $\mstar = 10^{8.5}\msun$, we
elect to apply scatter uniformly at all $\mpeak$.  Due to a lack
of galaxies in our observational sample with $\mstar\gtrsim10^9\msun$
and a lack of (sub)halos with $\mpeak\gtrsim10^{11}\msun$
(Figures~3~and~5 of \citealt{ELVIS}), however, our method primarily
constrains the scatter below the masses probed by \citetalias{Behroozi2013}.

Our second model, which we refer to as the ``growing scatter'' model,
is motivated by recent simulation results that suggest the level of
stochasticity in galaxy formation may increase at decreasing halo
masses \citep[e.g.][]{Sawala2014}.  Specifically, we model the log-normal
scatter $\sigma$ as a function of the peak mass, such that $\sigma = 0.2$
for $\mpeak > M_1$ but grows linearly with decreasing $\log_{10}\mpeak$
for $\mpeak \leq M_1$:
\begin{equation}
\sigma_\upsilon = 0.2 + \upsilon \times (\log_{10}\mpeak - \log_{10}M_1).
\label{eqn:growingsigma}
\end{equation}
The parameter $\upsilon$ quantifies the rate at which the scatter
grows:  a more negative $\upsilon$ (larger magnitude) results in
larger scatter at fixed $\mpeak$, whereas $\upsilon = 0$ corresponds
to a constant scatter of $0.2$~dex.

For both models, we create 500 realizations for each satellite or field
system, resulting in 12,000 ``paired'' realizations of the satellite
systems of the MW and M31 and 4500 realizations of the LF for every
combination of $\alpha$ and $\sigma$.

Though we do not present the results here, we also explore a model
with a sharp cut-off in star formation below $\mpeak = 10^9\msun$, as
motivated by reionization arguments \citep{Bullock2000,Somerville2002}
and recent hydrodynamic simulations (e.g. \citealt{Sawala2014}, but see
\citealt{Wheeler2015}).  The results from such a model are nearly identical,
however, as the galaxies included in our sample are nearly always assigned
to halos with $\mpeak > 10^9\msun$.  Moreover, the dark systems predicted
in \citet{Sawala2014} arise naturally from power-law models with scatter,
provided the scatter is large enough that (sub)halos can be assigned no
stellar mass. We additionally explore models with
one-sided scatter, including both exponential and Rayleigh distributions;
models with fixed $\sigma = 0.2$~dex above $M_1\sim10^{11.5}\msun$;
models that break from the \citetalias{Behroozi2013} relation at
$\mhalo = 10^{10.5}\msun$; and models with $\mstar$ fixed to be less than
$\mpeak \times f_\mathrm{b}$, the cosmic baryon fraction.  All of these
yield very similar results to the two simple models described above.
In principle, the $\mstar-\mhalo$ relation may not be well-described
by scatter about a power-law.  Without clear evidence of another functional
form, however, we elect to explore extrapolations of an AM relation that
reproduces the data at larger $\mstar$.

\begin{figure*}
\centering
\includegraphics[width=\columnwidth]{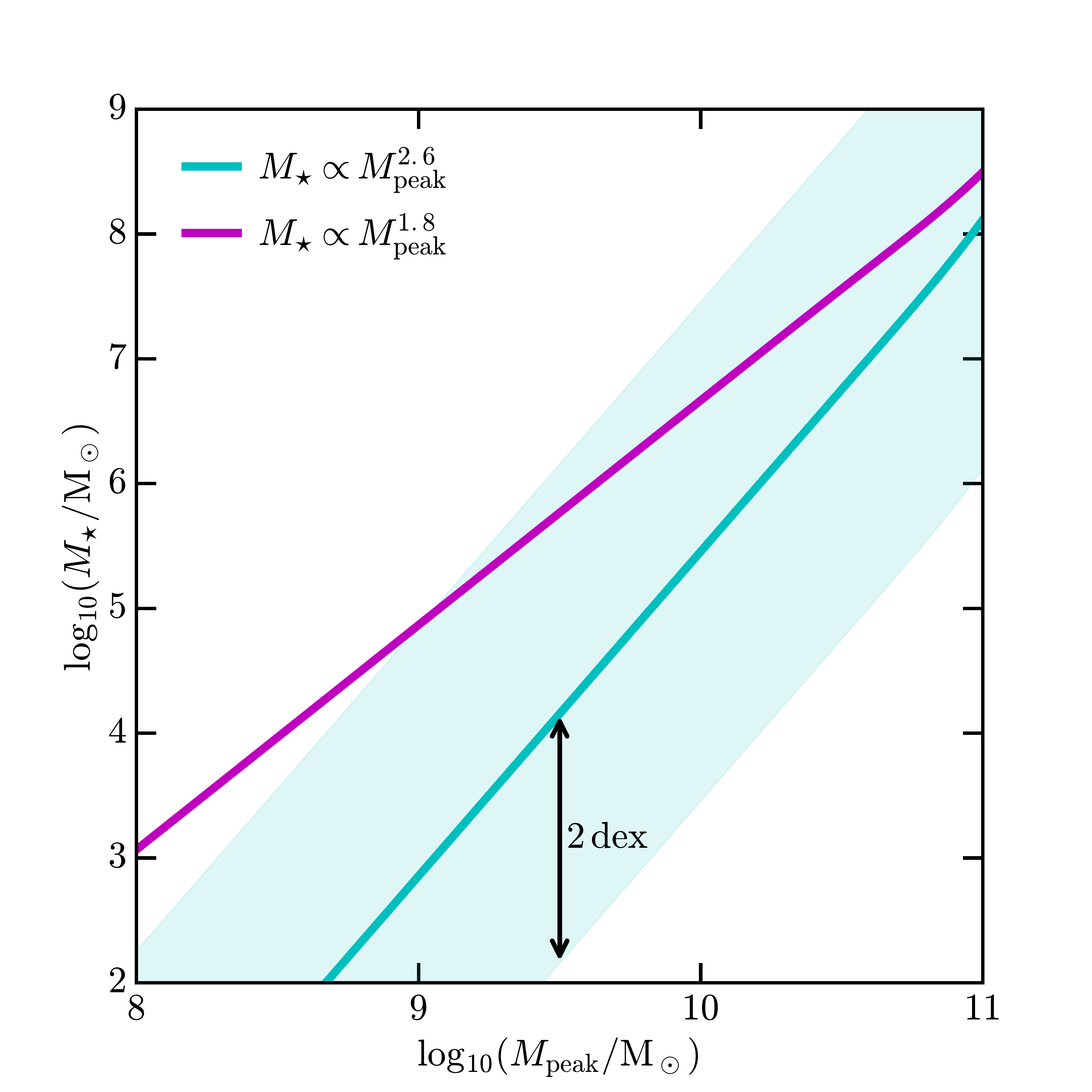}
\includegraphics[width=\columnwidth]{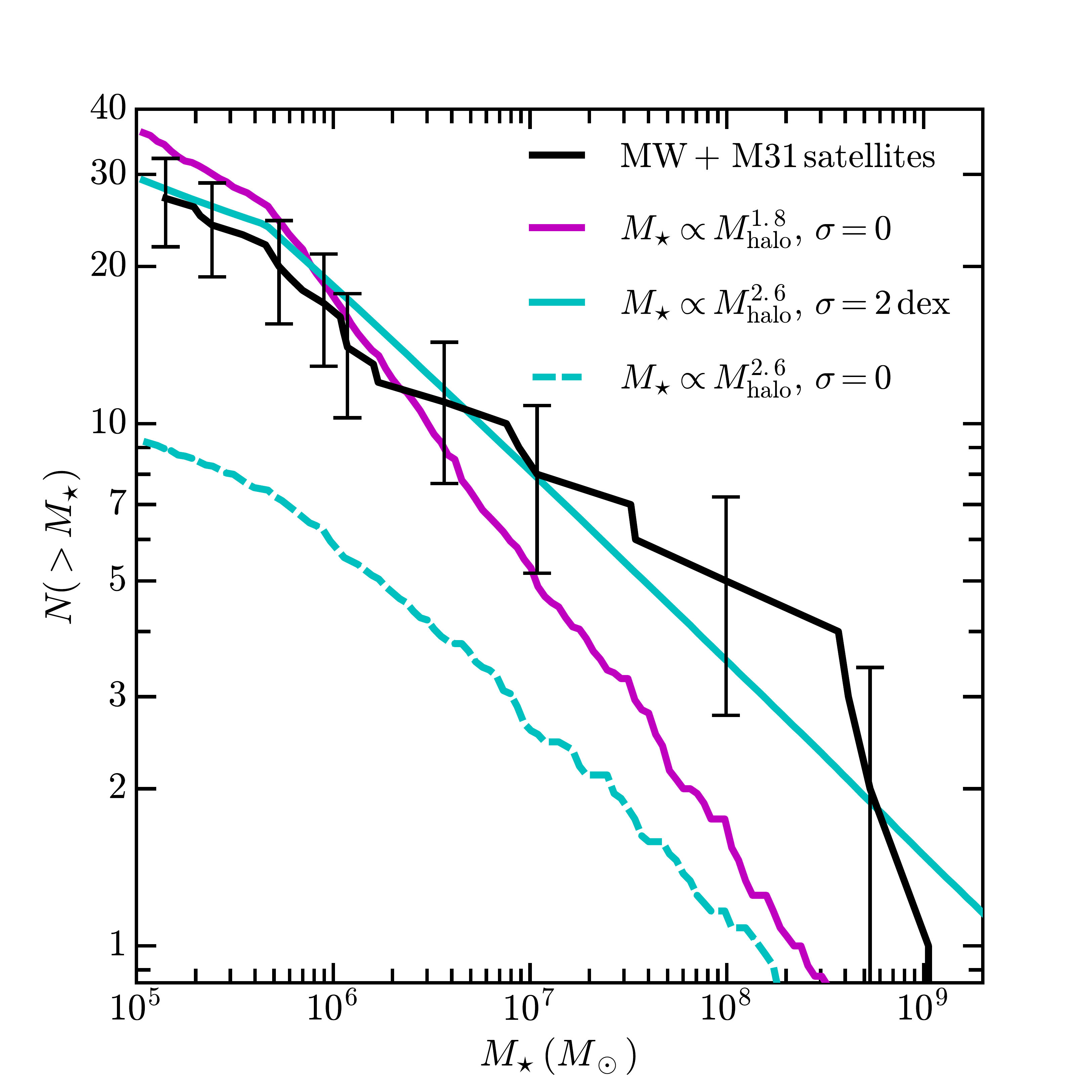}
\caption{\textit{Left:} Two sample relations between (sub)halo mass, quantified
here by $\mpeak$, the largest mass a (sub)halo ever attains, and galaxy stellar
mass $\mstar$ at $z = 0$.  Both relations are anchored to the \citet{Behroozi2013}
relation above $\mpeak\sim10^{11.5}$.  The magenta line plots a relation that at the
faint end asymptotes to $\mstar \propto \mhalo^{\alpha}$ with $\alpha = 1.8$
\citep[similar that advocated in][]{ELVIS} and with no scatter.  The cyan line
shows an alternative, steeper relation relation, with $\alpha = 2.6$, but with
correspondingly larger scatter $\sigma = 2$~dex.
\textit{Right:}  In black, the anti-cumulative stellar mass
function (SMF) of the MW and M31 satellites used in this work (see \S\ref{ssec:obs}), with
Poissonian error bars.  The magenta line plots the mean of the SMFs obtained
from applying the magenta line in the right plot to the ELVIS subhalo
populations.  Similarly, the solid cyan line plots the mean SMF obtained
by assuming the cyan relation in the right plot, including a log-normal
scatter of $2$~dex.  The dotted cyan line, however, represents the mean
SMF from the same relation, but without applying any scatter.
The break at $\mstar\sim10^{5.5}\msun$ is due to the completeness cut placed on
the MW satellites.  There is clearly a degeneracy between the slope of the
$\mstar$-$\mhalo$ relation and the assumed scatter, such that one must
vary both to maintain a good fit to the local SMFs.}
\label{fig:twoslopes}
\end{figure*}

\subsection{Quantifying the goodness of fit}
\label{ssec:kappa}

\begin{figure*}
\includegraphics[width=\columnwidth]{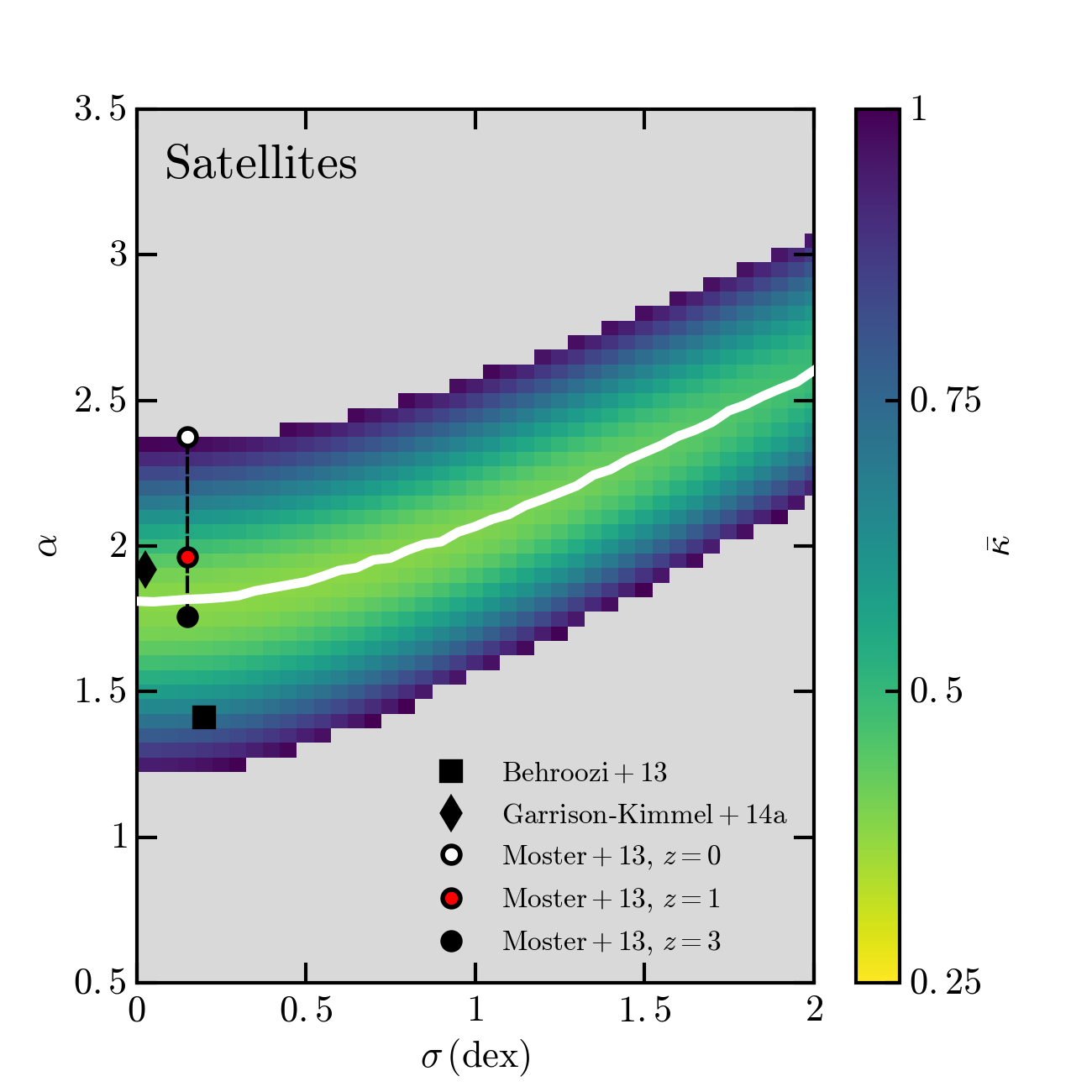}
\includegraphics[width=\columnwidth]{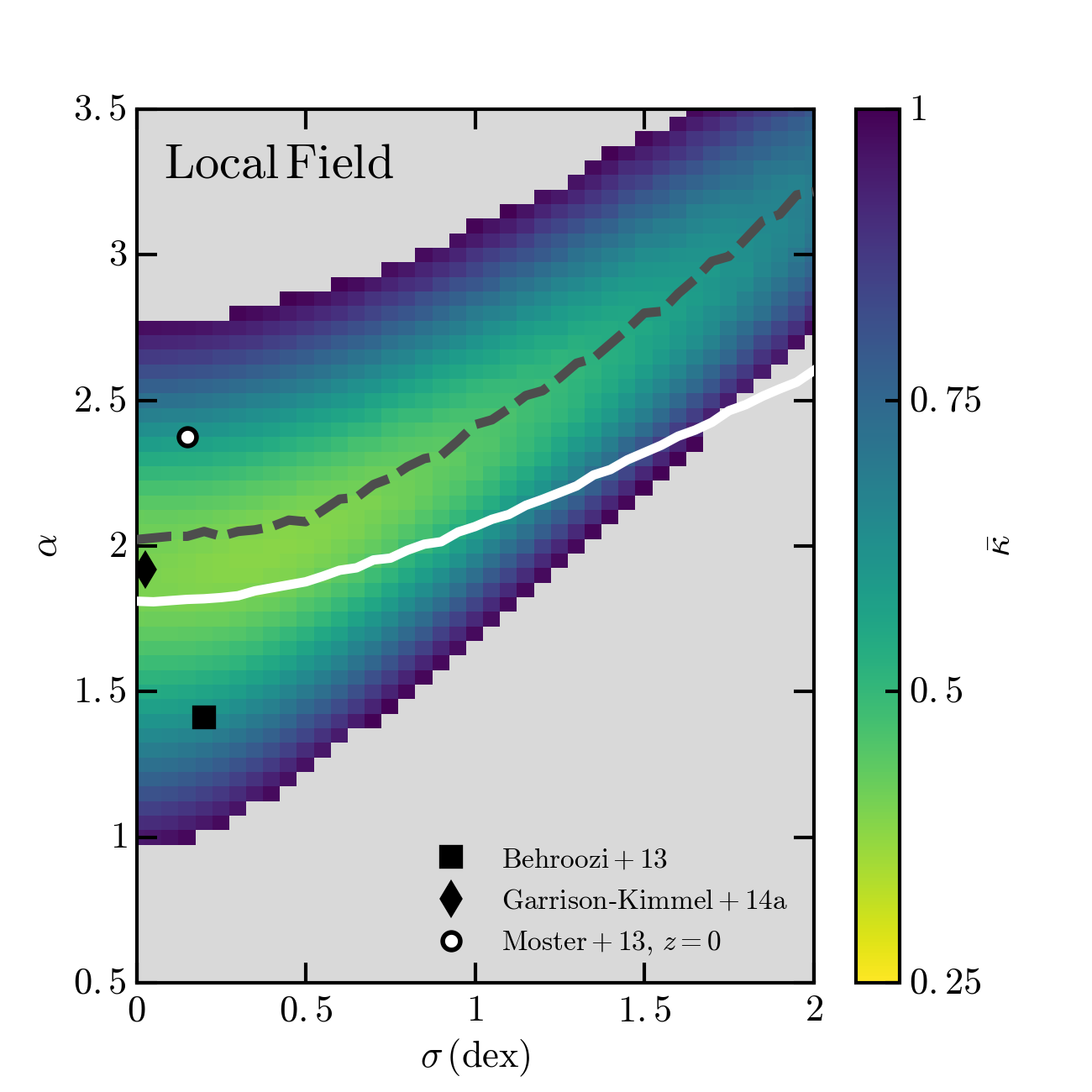}
\caption{The goodness of fit, quantified by $\kappa$, of the subhalo mass
functions to the MW and M31 stellar mass functions (left) and similarly for
halos and galaxies in the Local Field (right), as a function of the
log-slope $\alpha$ of the $\mstar$-$\mpeak$ relation and of the log-normal scatter
$\sigma$ about the median relation.  Plotted are the averages over the systems, which
are themselves averaged over 500 realizations for each combination of $\sigma$
and $\alpha$.  The solid white line, which is the same in both panels, indicates the
best-fit $\alpha$ derived from the satellite distribution.  For the left
plot, we minimize $\kappa_{\rm MW} + \kappa_{\rm M31}$ for each realization;
i.e., each host may be matched with either the MW or M31.  Plotted is
$(\kappa_{\rm MW} + \kappa_{\rm M31})/2$, which weights the two satellite systems
equally and consequently weights the individual MW satellites more heavily than
the M31 satellites.  As the assumed scatter increases, $\alpha$ must also increase
such that the relation falls off more steeply for $\mpeak < M_1\sim10^{11.5}\msun$.
Otherwise, the scattering of small halos to high $\mstar$ results in an over-estimation
of the SMF.  For comparison, we overplot the faint-end log-slopes quoted by
\citetalias{Behroozi2013} (black square), \citet[][black diamond]{ELVIS}, and
\citet[][circles]{Moster2013}.  Because the \citet{Moster2013} formalism dictates 
that satellite systems are assigned stellar masses based on both their infall mass 
and infall redshift, the left panel indicates the range of log-slopes quoted by 
\citet{Moster2013} for $0 < z < 3$.  We caution, however, that the \citet{Moster2013} 
and \citetalias{Behroozi2013} points represent extrapolations from higher stellar masses. 
We emphasize that a model that falls within the allowed band is not necessarily correct, 
but models that are purely power-laws for $\mstar\simeq10^5$ -- $10^8\msun$ that fall
outside the allowed band will not, in general, reproduce the LG $\mstar$
functions.}
\label{fig:kappagrids}
\end{figure*}

We now turn to the problem of quantifying the match between a theoretical
realization for the stellar mass function and the true observational SMF.
Rather than performing a likelihood-based analysis on the differential mass
functions, which is sensitive to the choice of bins due to the small numbers
of objects, we elect to compare the mass functions in an anti-cumulative sense.
Specifically, for each realization, we rank-order the galaxies by stellar mass
and calculate
\begin{equation}
\kappa = \frac{1}{N}\sum\limits_{i=1}^N \abs{\log_{10}\left(M_{\star,\,i}^{\rm theory}\right) - \log_{10}\left(M_{\star,\,i}^{\rm data}\right)}.
\end{equation}
That is, $\kappa$ for a single realization is the average difference
(in log-space) between the $i$-th brightest galaxy in the observational
sample and the $i$-th brightest galaxy in that realization.\footnote{We
also explore alternative definitions of $\kappa$, including a version
without the absolute value (such that differences may cancel out) and
a root-mean-square method, which highlights the largest discrepancies.
Overall, we find that the best-fit slopes are nearly identical between
the definitions when averaging over the realizations of a simulated
system, regardless of the specific definition of $\kappa$.}  The best-fit
log-slope for a given $\sigma$ is therefore that which minimizes $\kappa$.

In order to evaluate $\kappa$ as a metric, we compare the fiducial
stellar masses listed in Table~\ref{tab:obsdata} to Monte Carlo
realizations of the data themselves created by sampling log-normal
distributions on their reported observational errors in the stellar masses
(where the errors are again drawn from the data cataloged in
\citealt{McConnachie2012}).  The values of $\kappa$ calculated this way vary
between $\sim0.05$~dex and $\sim0.2$~dex, peaking at $\sim0.11$~dex, for both
the MW and M31 satellite systems.  The LF sample, by virtue of the high completeness
cut (and smaller errors on $\mstar$), peaks at $\sim0.06$~dex.  We conclude that
theoretical realizations with $\kappa \sim 0.2$~dex are very similar to the observed
stellar mass function, roughly consistent with randomly sampling the data themselves
within observational errors.

Because $\kappa$ takes into account only the $N$ ``brightest'' halos in a given
realization, where $N$ is the number of galaxies in the corresponding observational
sample, it does not constrain the shape of the stellar mass function below the
faintest galaxy in the observational sample.  This fact is inconsequential for the
MW and LF samples, as the incompleteness limit is defined by $\mstar$ of the
faintest galaxy, but it is important for M31, where the incompleteness is
better understood.  To therefore constrain the shape of the stellar mass function
between the incompleteness limit, $10^5\msun$, and the faintest galaxy in the sample,
$\sim1.5\times10^5\msun$, we insert a galaxy with $\mstar = 10^5\msun$ into the
catalog; that is, we assume there is an undetected galaxy lurking just below
the completeness limit.  Without this insertion, $\kappa$ generally prefers
slightly steeper slopes, but the models predict additional galaxies with
$\mstar\sim10^5\msun$ within the PAndAS footprint.\footnote{The number
of ``fake'' galaxies inserted, however, is relatively unimportant:  adding
ten galaxies with $\mstar=10^5\msun$ into the catalogs changes the results
by a negligible amount.}

\section{Results}
\label{sec:results}

One of the qualitative aims of this paper is to accurately explore the
covariance between scatter and slope.  The basic need for this is illustrated in
Figure~\ref{fig:twoslopes}.  The left panel shows two example relations between stellar
mass and halo mass: one, in cyan, with a steep slope
($\mstar \propto \mhalo^\alpha$ with $\alpha = 2.6$) and large scatter ($\sigma = 2$ dex),
and the other, in magenta, with a flatter slope ($\alpha = 1.8$) and no scatter.  The right
panel presents the resultant stellar mass functions from averaging over realizations of the
ELVIS halos, along with those observed for the MW and M31 combined (see
\S\ref{ssec:obs} and Table~\ref{tab:obsdata} for details).  Both of these models
produce similar stellar mass functions (which are roughly consistent with those observed
in the Local Group) even though their slopes are significantly different.  This is driven by
the difference in scatter.  The dashed cyan line the right panel assumes the same steep relation
as the solid line, but this time with zero scatter; this results in a satellite mass function that
is much below the data. This difference is due to the rapidly rising halo mass function.  In cases
with a large amount of scatter, small systems scattering to large stellar masses will dominate over
large systems scattering to low stellar masses.  Intuitively, therefore, large scatter requires a
steeper slope to avoid overproducing the observed stellar mass function.

In what follows we present a more rigorous analysis to jointly constraint the scatter and slope.

\subsection{Relation between log-slope and scatter}
\label{ssec:sigmaalpha}

We begin by examining the best-fit log-slope $\alpha$ at each assumed
scatter $\sigma$, both within the virialized volumes of the host halos
and in the fields surrounding them.
Figure~\ref{fig:kappagrids} shows the fit parameter $\kappa$ as a function of
slope $\alpha$ and scatter for the satellites of the MW and M31 on the left and
galaxies in the Local Field (LF) on the right.  The results have have been averaged
over all simulations, including 500 realizations of each simulation for each
($\alpha$, $\sigma$) model represented on the grid. In the left panel, we choose the
smaller of $(\kappa_{\rm MW,\,host\,1} + \kappa_{\rm M31,\,host\,2})/2$
and $(\kappa_{\rm MW,\,host\,2} + \kappa_{\rm M31,\,host\,1})/2$
for each realization, intrinsically allowing the MW/M31 to be matched
with either the larger or smaller ELVIS host.\footnote{Only one ``pair''
strongly prefers to match the MW with the larger host at all $\sigma$,
whereas several prefer to match the MW with the lower mass host.
However, for $\sigma\gtrsim0.25$, many systems have roughly an
equal probability to match the MW with each host.  Together, these
conspire to keep the mean fraction of realizations where
the MW is matched with the larger host nearly constant at $\sim30\%$.}

The solid white line in Figure~\ref{fig:kappagrids} indicates the average
best-fit $\alpha$ for the satellite systems; i.e., it plots the minimum $\kappa$
in the left panel.  The two regions yield qualitatively similar results, though
the LF prefers a slightly steeper slope (indicated by the dashed grey line in the
right panel of Figure~\ref{fig:kappagrids}) and marginally prefers a smaller $\sigma$,
whereas the satellite analysis yields an equally good fit at all $\sigma$.
In practice, the two regions may be combined to yield a single
$\alpha(\sigma)$ relation, but we elect to consider them independently,
both to include the isolated ``pairs'' in the left panel, which do
not have analogous LF regions and therefore cannot be included in the right
plot, and because the incompleteness is more poorly understood in the LF.
The average best-fit $\alpha$ is well fit by a quadratic function in both
cases.  Specifically, for the satellites we find $\alpha$:
\begin{equation}
\alpha_{\rm sats} \cong 0.14 \sigma^2 + 0.14 \sigma + 1.79
\label{eqn:basats}
\end{equation}
while the LF relation is best-fit by
\begin{equation}
\alpha_{\rm LF} \cong 0.24 \sigma^2 + 0.16 \sigma + 1.99,
\end{equation}
where $\sigma$ is one standard deviation of the $\log_{10}$-normal scatter.

The symbols overplotted in Figure~\ref{fig:kappagrids} indicate the results
of \citetalias{Behroozi2013} (black square), \citet[][black diamond]{ELVIS},
and \citet[][circles]{Moster2013}.  Because the latter relation is derived
assuming that abundance matching is performed at the redshift of a subhalo's
infall, we plot the evolution of the slope for $0 < z < 3$ in the left panel
(dashed black line), with specific redshifts highlighted.  Though there is a
large spread in the infall redshifts of the ELVIS subhalos, \citet{Wetzel2015}
showed that most fell into a larger host between $3 \lesssim z \lesssim 0.5$.
Nonetheless, the comparison to the \citet{Moster2013} slopes is imperfect due to
differing normalizations; in practice, however, the \citet{Moster2013}
normalization at $\mhalo = 10^{11}\msun$ differs from that of \citetalias{Behroozi2013}
(and therefore our normalization) by less than $5\%$ in logspace.

Figure~\ref{fig:growinggrid} similarly illustrates the results for a model where
the scatter grows with decreasing galaxy mass.
As in Figure~\ref{fig:kappagrids}, the vertical axis shows the
log-slope of $\mstar-\mpeak$, but the horizontal axis here plots $\upsilon$,
the slope of $\sigma(\log_{10}\mpeak)$ (see Equation~\ref{eqn:growingsigma}).
For reference, the upper axis indicates the symmetric scatter at
$\mpeak = 10^{10}\msun$, highlighting the fact that the growing
scatter model explores larger scatters than the constant scatter
case.  The results, however, are qualitatively identical:
as the scatter grows, $\alpha$ must also grow in order to avoid overproducing
the observed SMF.  The best-fit relation is here presented in terms of $\upsilon$:
\begin{equation}
\alpha_{\upsilon,{\rm sats}}  \cong 0.25 \upsilon^2 - 1.37 \upsilon + 1.69.
\end{equation}
We do not plot the relation for the LF because it follows a qualitatively
identical trend to Figure~\ref{fig:growinggrid} with a slight offset to
steeper slopes, as expected through comparison with Figure~\ref{fig:kappagrids}.
Specifically, the best-fit log-slope $\alpha$ from the field galaxies is given by
\begin{equation}
\alpha_{\upsilon,{\rm field}} \cong 0.47 \upsilon^2 - 1.48 \upsilon + 1.81.
\end{equation}

The difference in the preferred log-slopes between the satellite systems
and the LF is likely to be enhanced by tidal effects from a baryonic disk,
which will systematically shift the subhalo mass functions lower at fixed
counts.  Assuming that dark subhalos are destroyed at the same rate as
luminous satellites, this shift will allow an even shallower $\mstar-\mhalo$
relation to fit the satellite systems.  The discrepancy between the
LF and the satellite systems is further highlighted by extrapolating the
the LF below the completeness limit:  the best-fit relation
at $\sigma = 0$~dex predicts an average of $\sim40$ LF galaxies with
$10^5 < \mstar/\msun < 10^6$ (``classical'' dwarfs), which decreases to
$\sim10$ for $\sigma = 2$~dex.  If upcoming surveys do not identify a wealth
of new nearby field galaxies, it may be indicative of a change in how
galaxy formation proceeds in small halos at late times, as satellites of a
given mass today are likely to have formed earlier than field halos at the same
mass.

Averaged over the twenty-four MW-M31 pairs or the nine LFs, the best-fit
$\alpha$ yields a fit, quantified by $\kappa$, that is $3-4$ times worse
than the Monte Carlo realizations of the observational data with themselves.
However, the single best-fit \emph{realization} of each system, taken over
all values of $\alpha$ at fixed $\sigma$, typically yield a fit that is only
$\sim0.1$~dex worse than the data with themselves, even after averaging over
the systems.  Moreover, merely selecting these best-fit realizations at each
$\sigma$ yields log-slopes identical to the fits given above.

\subsection{Implications for the too-big-to-fail problem}
\label{ssec:tbtf}
Having quantified the best-fit log-slope as a function of the scatter,
we move to investigating how the severity of TBTF varies with $\sigma$.
We quantify the problem by counting the number of massive dark matter
halos that are missing from the known galaxy population. We call these
missing massive systems ``massive failures."
While Figure~\ref{fig:kappagrids} indicates the trends with $\alpha$ and
$\sigma$ averaged over the ELVIS systems, in this section we will count massive
failures for a given $\sigma$ at the best-fit $\alpha$ \emph{of each system}, averaging over
the realizations of that system.  Within $300$~kpc of the MW, we then count
massive failures in the host matched with the MW, again in a
realization-by-realization manner.

Our definition of massive failures is similar to that employed
in \citet{ELVISTBTF} in that we select ``candidate'' (sub)halos to
host the dwarf galaxies, then require that each dwarf be matched
with only one (sub)halo.  A dwarf may only be matched with a (sub)halo
if the $1\sigma$ error on the implied circular velocity at the 3D
half-light radius, $\vhalf$ at $\rhalf$, is greater than the circular
velocity $\vcirc$ of that sub(halo) at $\rhalf$.  As in \citet{ELVISTBTF},
we calculate $\vcirc$ profiles for halos from $\vmax$, the maximum circular
velocity of the halo, and $\rmax$, the radius at which $\vmax$ occurs,
assuming \citet{Einasto} profiles with a shape parameter of $0.18$, which
fits the density profiles of subhalos in ultra-high resolution simulations slightly
better than the \citet{NFW} profile \citep{Springel2008}.\footnote{As
demonstrated in \citet{ELVISTBTF}, using the slightly lower density NFW profile reduces
massive failure counts around the MW by $\sim$15\%, but has a negligible effect
in the Local Field.}
Unlike \citet{ELVISTBTF}, however, we select candidate halos based not on
their halo properties, but on the stellar masses that they were assigned
in a given realization, such that all halos above the incompleteness limits
($4.5\times10^5\msun$ around the MW and $4.5\times10^6\msun$ in the LF)
are eligible to be counted as massive failures.  All remaining (sub)halos,
i.e. those with $\mstar$ above the completeness limit that are not
assigned a galaxy, are considered massive failures.  The average number
of these massive failures in each realization is plotted in
Figure~\ref{fig:tbtf}, which we explain in greater detail below.

Our observational samples are comprised of the MW satellites
and LF dwarfs listed in Table~\ref{tab:obsdata}.  As in previous
TBTF analyses \citep[e.g.][]{TBTF2,ELVISTBTF}, however, we eliminate
the LMC and the SMC due both to their relative rarity and substantial
contribution from baryons within $\rhalf$, and also exclude the
Sagittarius dSph as it is currently interacting with the disk and
is unlikely to be in kinematic equilibrium.  For consistency, we
therefore rank order the subhalos of each realization by their
assigned $\mstar$ and exclude the three largest, rather than simply
excluding the three most massive (by dark matter mass).  We also
exclude NGC~6822 from our counting in the LF, as it is likewise baryon
dominated within $\rhalf$ \citep{Kirby2014}, and Sagittarius~dIrr, as
there have thus far been no attempts to measure its internal kinematics.
To match these exclusions, we similarly remove the halo with the highest
$\mstar$ and the halo with the sixth highest $\mstar$ from each realization
for the LF.

Kinematic data ($\vhalf$ and $\rhalf$) for the remaining nine MW
satellites are drawn from \citet{Wolf2010}, who used observational
data from \citet{Munoz2005,Koch2007,Simon2007,Mateo2008} and
\citet{Walker2009}.  We similarly apply the \citet{Wolf2010} formula
to measurements taken by \citet{Kirby2014} for the purely dispersion
supported galaxies in the LF.  However, two of the LF galaxies included
display evidence of additional rotational support.  $\vhalf$ and $\rhalf$
for WLM are taken from \citep{Leaman2012}, and, following
\citet{Weiner2006}, we account for the rotation support of Pegasus
dIrr by adding,  in quadrature, the projected rotation velocity
$v\sin i$ and the stellar velocity dispersion $\sigma_{\rm star}$
when using the \citet{Wolf2010} formula for $\vhalf$
\citep[also see \S5.2 of][]{Kirby2014}.

\begin{figure}
\centering
\includegraphics[width=\columnwidth]{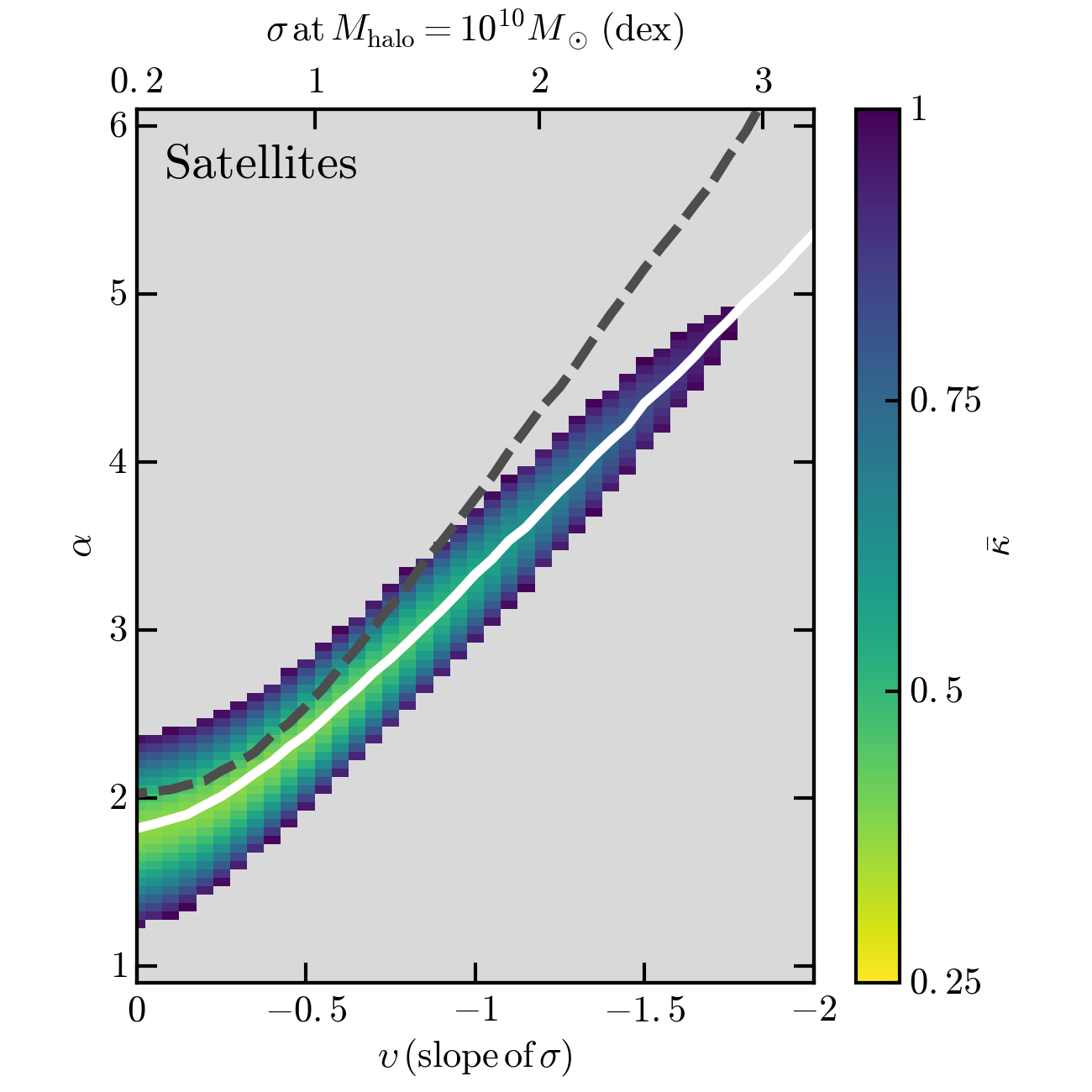}
\caption{The goodness of fit between the MW/M31 SMFs and the mock satellite
catalogs obtained by applying the growing scatter model to the ELVIS
simulations.  As in Figure~\ref{fig:kappagrids}, the y-axis indicates the
assumed faint-end log-slope $\alpha$.  Here, however, the x-axis indicates
$\upsilon$, the log-slope of $\sigma$ (see Equation~\ref{eqn:growingsigma}).
For reference, the upper x-axis shows the resultant scatter at
$\mpeak = 10^{10}\msun$.  The x-axis is inverted such that the scatter
increases towards the right.  As in the constant scatter model, a larger
$\sigma$ necessitates a more rapid fall-off in $\mstar(\mpeak)$ to avoid
overproducing the observed SMFs.  Because we explore even larger $\sigma$
compared to the constant scatter model, however, we find even steeper
slopes are required.  As in Figure~\ref{fig:kappagrids}, the solid white
and dashed grey lines represent the best-fit models in for the satellite
and field systems, respectively.}
\label{fig:growinggrid}
\end{figure}

\begin{figure*}
\includegraphics[width=\columnwidth]{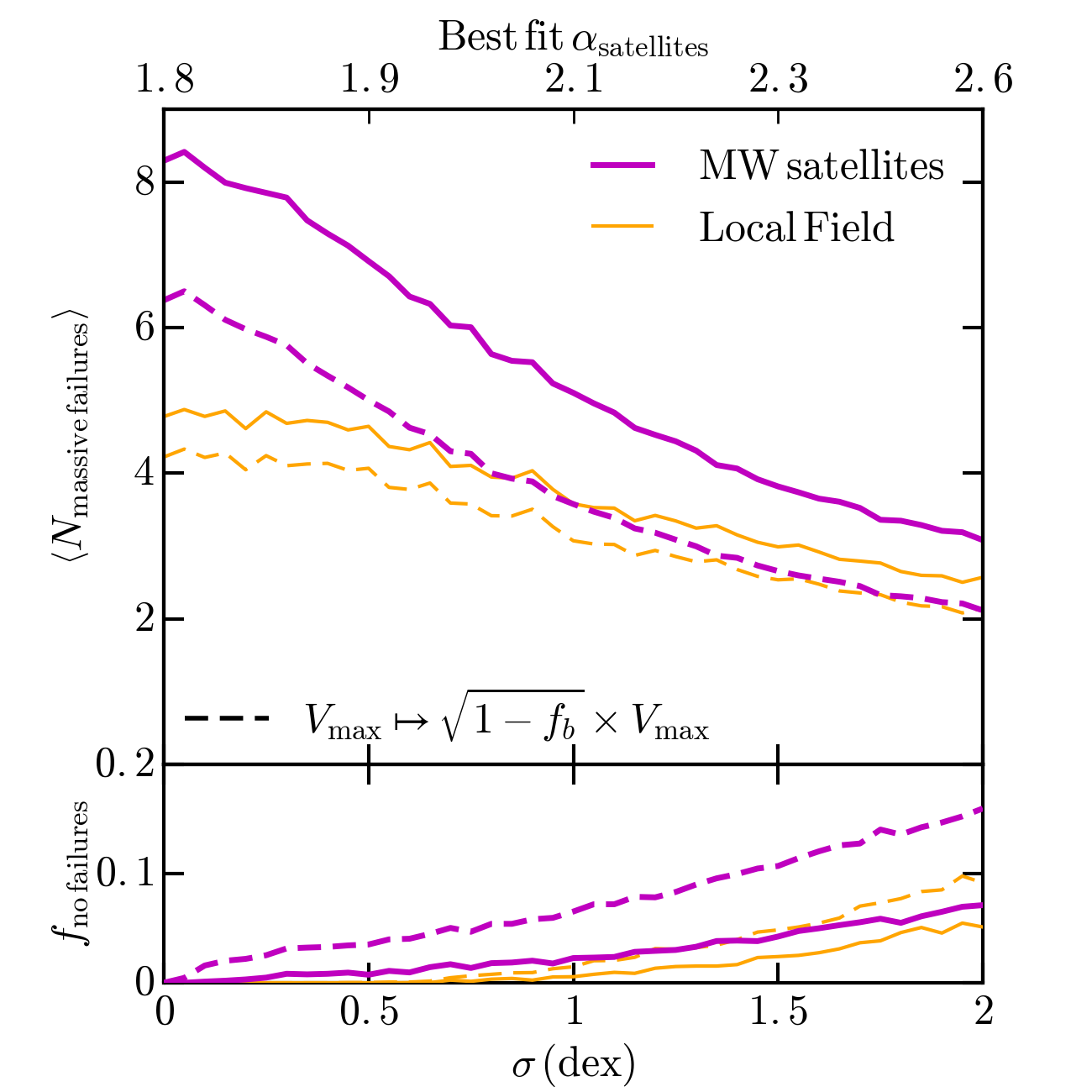}
\includegraphics[width=\columnwidth]{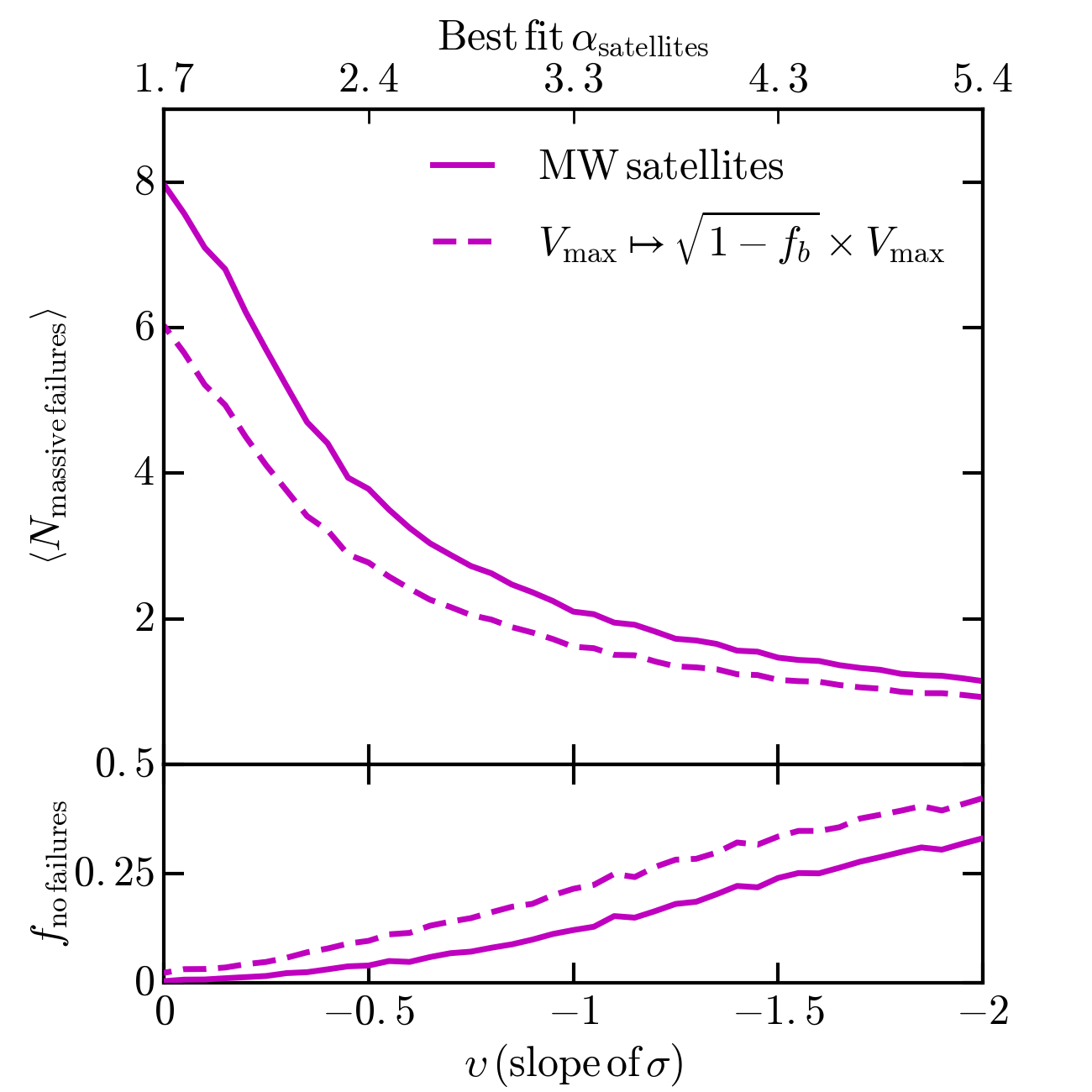}
\caption{The number of massive failures, within $300$~kpc of the MW
in magenta and in the Local Field in green, as a function of the
scatter at the best-fit $\alpha$ at the scatter in the constant scatter
model (left) and the growing scatter model (right), plotted such that
the magnitude of the scatter increases towards the right of each plot.
Massive failures are defined as (sub)halos with $\mstar$ greater than
the completeness limit ($4.5\times10^5\msun$ in the MW, $4.49\times10^6\msun$
in the Local Field) without observational counterparts (as in
\citealt{ELVISTBTF}), where the three brightest (by $\mstar$) subhhalos and
the brightest and sixth brightest field halo in each realization are excluded,
to match the observational cuts.  The magenta lines plot the number of
massive failures within $300$~kpc of the MW, averaged over all $12$ pairs
and the $12$ pseudo-pairs; the orange line averages over the fields
of the nine pairs included in our analysis. For each realization,
massive failures within $300$~kpc are drawn from the subhalos of the host
matched with the MW.  The dashed lines indicate massive failure counts after
suppressing the rotation curves of all halos by $\sqrt{1-f_b} \sim 0.91$ to
mock the effects of a reduced particle mass in cosmological hydrodynamical
simulations.  As the scatter increases, the average number of massive
failures within 300~kpc in the constant scatter model (growing scatter model)
decreases by $\sim2/3$ ($\sim7/8)$, and the fraction of realizations with no
massive failures increases to $\sim10\%$ ($\sim40\%$).  The field demonstrates
a similar trend, such that $\sim5\%$ of realizations are without massive
failures in the constant scatter model.}
\label{fig:tbtf}
\end{figure*}

The results of this counting are summarized in the left and right
panels of Figure~\ref{fig:tbtf} for the constant and growing scatter
models, respectively.  The upper panels plot the number of massive
failures, either within $300$~kpc of the MW (magenta lines) or in
the LF (orange lines), as a function of the assumed scatter $\sigma$
(left panel) or the assumed log-slope of the scatter $\upsilon$ (right panel),
while the lower panels indicate the fraction of realizations with
zero massive failures.  Solid lines indicate the fiducial count,
roughly comparable to \citet{ELVISTBTF}, while the dashed lines
count massive failures after suppressing the $\vmax$ of every halo by
$1 - \sqrt{1-f_b} \sim 10\%$ to demonstrate a scenario wherein
the baryonic contribution to the particle mass in the DM-only ELVIS
simulations is removed from the potential, thereby partially capturing
that dwarfs are typically far from baryonically closed \citep[e.g.][]{Wolf2010}.
As in Figures~\ref{fig:kappagrids}~and~\ref{fig:growinggrid}, the lines
plot averages over the systems, which are themselves averaged over the
realizations of that system.  We do not count massive failures around
the host matched with M31, but note that \citet{Tollerud2014} showed
that TBTF is similarly extant among the M31 satellites.

The results of Figure~\ref{fig:tbtf} are striking, particularly
for the satellites of the MW.  On average, the number of massive failures
decreases from $\sim 8$ and $\sim 6$ in the zero-scatter count, for the standard
and gas-loss cases, down to $\sim 2-3$ for both cases when  $\sigma$
increases to 2~dex.\footnote{We note that there are significantly fewer massive
failures counted here than in \citet{ELVISTBTF}, who found $\sim15$
massive failures in each LF.  This is primarily due to the more
stringent cuts that we have used in this work: only halos assigned
enough stellar mass in a given realization to be above the
completeness cut are eligible to be failures.  In the previous work,
we selected halos with peak $\vmax > 30~\kms$.} Moreover, as  seen in
the bottom panels, the fraction of realizations that are
completely devoid of massive failures increases steadily with increasing scatter.
For 2 dex of scatter, we find that  $\sim 15\%$ of realizations are completely
without massive failures, once we account for the reduction in halo mass due
strictly to the reduced mass of DM particles in cosmological hydrodynamical
simulations.  In the right-hand panel we see that the growing scatter model
yields similar results.

Our method is designed to maintain a roughly equal number of halos
(9 in the MW and 7 in the LF) with stellar masses large enough to be seen above
the adopted completeness limits at all
$\sigma$ and $\upsilon$.  This means that for  low scatters (1) even the
best-fit $\alpha$ overproduces the stellar mass function of
the MW at low masses (see Figure~\ref{fig:twoslopes}) and (2) the vast
majority ($\sim 60\%$ in the MW sample and $\sim 85\%$ in the LF) of
halos above the completeness limit
are massive failures.  At high scatters, however, a larger portion
of the theoretical sample is composed of small halos (hosting bright
galaxies) that are more kinematically similar to the known dwarfs.

\begin{figure*}
\centering
\includegraphics[width=\columnwidth]{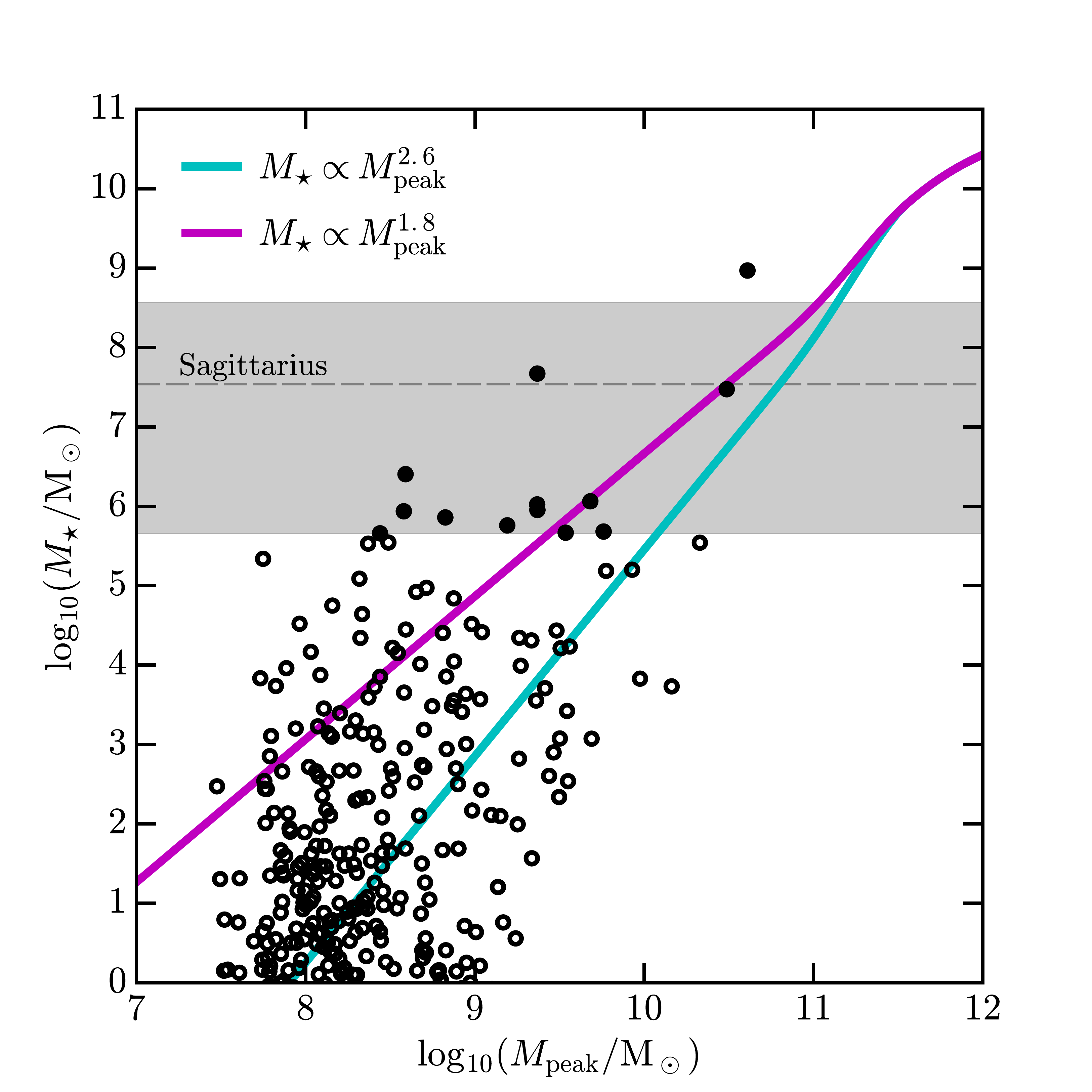}
\includegraphics[width=\columnwidth]{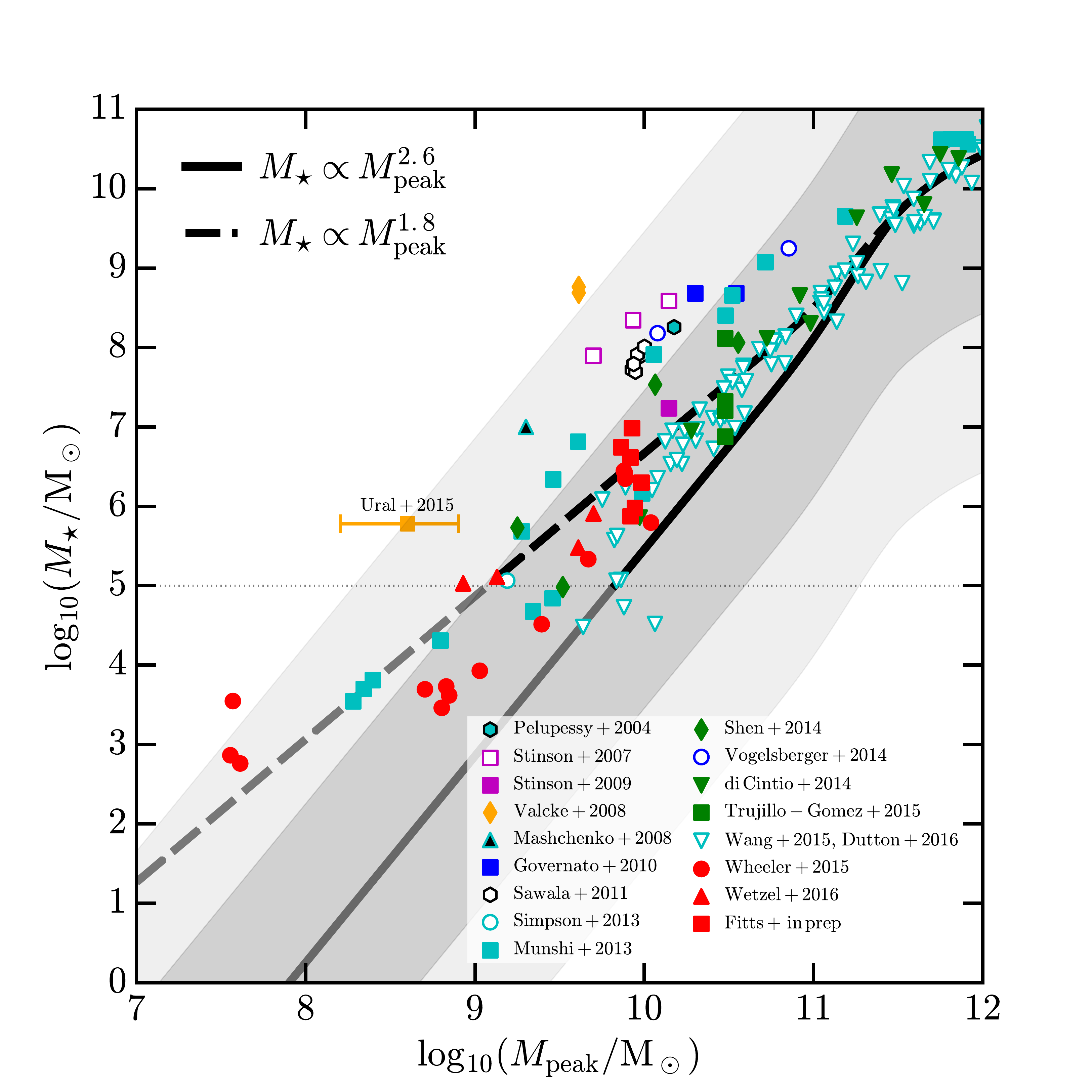}
\caption{\textit{Left:}  A sample realization of the MW satellite system,
obtained by applying the best-fit $\sigma=2$~dex relation (black circles);
the median relation is indicated by the cyan line.  For comparison, the
magenta line plots the best-fit zero-scatter relation.   The shaded
region roughly indicates the range in stellar masses probed by our method
in the MW (constraints from M31 extend to $10^5\msun$).  The upper edge of this
regime is here set by the stellar mass of the SMC, but in general is defined by
the second highest $\mstar$ in each realization.  Subhalos that fall below
the completeness limit are indicated by open circles.  The horizontal dashed
line indicates the stellar mass of Sagittarius, which roughly sets the upper
limit on the massive failure sample (the exact limit is set by the
third most luminous subhalo in each realization; see \S~\ref{ssec:tbtf});
subhalos that fall between this limit and the lower edge of the shaded band
are eligible to be massive failures.  The 2~dex case, selected because it
has no massive failures, illustrates how scatter in $\mstar-\mhalo$ is able
to alleviate TBTF:  many of the most massive subhalos, which have internal
kinematics inconsistent with the MW dSphs, are assigned to host ultra-faint
dwarf galaxies that lie in the incomplete regime, and are therefore excluded
from the analysis.
\textit{Right:}  A comparison of the same two best-fit relations (black lines,
which become gray when extrapolated beyond the regime constrained by our analysis
in either the MW or M31, $\mstar < 10^5\msun$, a limit that is also
indicated by the horizontal dashed line) to a sample of simulated dwarf
galaxies in the literature (references are as indicated in the legend); the shaded
regions indicate one and two standard deviations. If the simulations trace the
best-fit zero-scatter relation (dashed line), then the level of scatter exhibited
overall will generally overproduce the local SMFs.  Alternatively, if the simulations
trace the upward scatter about the $2$~dex relation, which aids in resolving TBTF,
then they thus far fail to sample the downward scatter necessary to avoid
overproducing the local SMFs.  Taken independently, many of the simulation
suites targeting isolated dwarfs, including \citet{Munshi2013}, NIHAO
\citep{Wang2015,Dutton2016}, and those simulated with the FIRE physics
\citep{Hopkins2014}:  \citet{Wheeler2015,Wetzel2016}, and Fitts et al. (in preparation),
may be sampling a relatively low-scatter relation (with $\sigma <~ 0.5$~dex) that
would generally reproduce LG counts.
However, a larger sample of halos, particularly with $\mpeak < 10^{10} \msun$,
is necessary to make quantitative statements about the behavior of simulations
in this regime, where constraints remain weak relative to higher $\mstar$.
The \citet{Ural2015} point (orange square with error bars),
which is derived from combining N-body simulations with MCMC modeling,
represents a potential constraint on Carina and is discussed in \S\ref{sec:discussion}.}
\label{fig:nofails}
\end{figure*}

Relations with very large scatter solve TBTF by occasionally assigning the
largest halos a very small stellar mass~--~so small that they
fall below the completeness limit.   Effectively, these relations solve
TBTF by demanding that the most massive halos do indeed fail to form stars
appreciably.    The $\mstar-\mhalo$ relation for one such realization
($\sigma = 2$~dex) is plotted explicitly in the left panel of
Figure~\ref{fig:nofails} as black circles; the median relation is indicated
by the cyan line.  The open points plot subhalos that lie below the
completeness limit of the MW (indicated by the lower edge of the shaded band),
and which are therefore ineligible to be massive failures.  This realization
does not exhibit the TBTF problem because many of the largest subhalos are
assigned to host ultra-faint dwarf galaxies, which we assume are unobservable.
By contrast, the best-fit zero-scatter relation (magenta line) predicts that
multiple high-mass subhalos host galaxies that lie above the completeness cut,
and would therefore be counted as massive failures.

The right panel of Figure~\ref{fig:nofails} shows the two $\mstar-\mhalo$ models
presented in the left panel along with a summary of recent predictions from
high-resolution hydrodynamic simulations. The black lines, which are extrapolated
as gray lines below the dotted horizontal line at $\mstar = 10^5\msun$,
the smallest $\mstar$ cut adopted, again represent the best-fit models for
$\sigma = 0$ (dashed line) and $\sigma = 2$~dex (solid line, with shading to indicate
$\sigma$ and $2\sigma$), but the points listed in the lower-left indicate the
results of baryonic simulations that include both star formation and feedback
\citep[][Fitts et al. in preparation]{Pelupessy2004,Stinson2007,Valcke2008,Mashchenko2008,Stinson2009,Governato2010,Sawala2011,Simpson2013,Munshi2013,Shen2014,Vogelsberger2014,diCintio2014,Trujillo-Gomez2015,Wang2015,Dutton2016,Wheeler2015,Wetzel2016}.

Though the simulations adopt a range of baryonic prescriptions, it is
immediately obvious that, taken together, they qualitatively fail to
uniformly sample the $\sigma = 2$~dex relation.  Specifically, only the
NIHAO simulations form low-mass galaxies in $\mpeak~10^{10}\msun$ hosts,
which is what is required to solve TBTF via scatter alone.  Alternatively, if
the simulations trace a flatter relation, then the scatter about that relation
is too large to avoid overproducing the SMFs with low-mass halos hosting bright dwarfs.
We caution, however, that much of the apparent scatter is due to variations
among different codes.  Many of the individual simulations produce relations
that are rather tight on their own, and several may reproduce LG galaxy counts
if applied to an LG-like volume.  However, more high-resolution simulations targeting
halos with $\mpeak \lesssim 10^{10}\msun$ are needed to fully quantify the behavior of
any of these codes in this regime.

\section{Discussion}
\label{sec:discussion}
Our results highlight the importance of understanding the
nature of the $\mstar$-$\mhalo$ relation at low masses.  They
also indicate that, if the $\mstar-\mhalo$ relation behaves
as a power-law with scatter at low $\mstar$, pinpointing the
degree of either the scatter or the slope of the median relation
should provide clues to the other, though it may be impossible to
disentangle the two.  Moreover, the level of scatter has clear
consequences on current small-scale problems in the $\Lambda$ 
cold dark matter ($\Lambda$CDM) paradigm of galaxy formation.  
Specifically, a high scatter results in a significant fraction of 
satellite and LF systems wherein the majority of the massive 
($\mhalo\sim10^{10.5}\msun$) halos preferentially scatter downward in 
stellar mass, thereby alleviating TBTF by assigning the typically problematic 
halos to host ultra-faint dwarf galaxies. This requires a steep fall-off 
in the median relation, however. Furthermore, the required scatter is 
very high:  for $\sigma = 2$~dex, $5\%$ of (sub)halos will host galaxies 
that lie a factor of $10^4$ from the median relation.

Ultimately, however, it is extremely difficult to determine
whether the rapid variability of the instantaneous specific
star formation rate predicted in simulations \citep[e.g.][]{Onorbe2015,Wetzel2016}
and the sensitivity to mass accretion discussed in \S~\ref{sec:intro}
truly does result in $\gtrsim2$~dex of scatter in $\mstar$ at fixed
$\mhalo$ in the dwarf regime.  A conclusive answer would require
not only an enormous sample of dwarf galaxies in the field,
where DM halos have not been stripped by larger hosts, but
measurements of the \emph{halo} masses of each of those dwarfs.
It would therefore be necessary to trace the mass well beyond
the stars, or to firmly understand the density profiles
of dwarfs such that measurements on small radii can accurately
be extrapolated to the virial radius.

Recently, \citet{Ural2015} demonstrated an alternative
approach by running idealized N-body simulations to model Carina
($\mstar = 6.03\times10^5\msun$), wrapped within an MCMC pipeline
to marginalize over the parameters of the initial conditions, including
the pre-infall (peak) halo mass.  Applying this method to a large number
of MW and M31 satellites could potentially provide valuable insights into
the shape and scatter of $\mstar-\mhalo$.  Their result, which is plotted
in Figure~\ref{fig:nofails} as an orange square and which agrees well with
results of likelihood-based comparisons with high-resolution cosmological
simulations \citep{TBTF2}, places Carina well above the best-fit $\sigma=0$
relation.  It may therefore be a $\sim2$ standard deviation outlier from the
$\sigma=2$ relationship.

Large scatter in $\mstar$-$\mhalo$ may also be
inferred from the properties of known galaxies in the MW.
For example, \citet{Wheeler2015} identified a mass-dependent
cutoff in post-reionization star formation in cosmological
simulations:  only halos larger than $\sim5\times10^9\msun$
are able to retain their gas in the presence of a reionizing
background and continue to form stars beyond $z\sim6$.  Taken
together with the results of \citet{Brown2014}, who showed that
six of the MW ultrafaint ($\mstar\sim10^4\msun$) satellites
are dominated by stars older than $12$~Gyr, one may easily
conclude that the ultrafaint galaxies live in halos smaller
than $\mhalo\sim5\times10^9\msun$.  A large scatter about the
median relation, however, would result in several ultra-faints
residing in large dwarfs.  This scatter may therefore manifest
in the form of ultra-faint dwarf galaxies with extended star
formation histories.  As was recently shown by \citet{Monelli2016},
Andromeda XVI may be one such system.  We may also expect to
find faint, dense galaxies living at the centers of halos too
massive to be altered by the small number of feedback events
from those galaxies.

Our conclusions are largely similar to those of \citet{Guo2015}, who
applied semi-analytic galaxy formation models to DM-only simulations,
then selected hosts analogous to the MW based on the observable
properties of the system.  Their models predict highly stochastic
galaxy formation at small mass scales, yielding a large scatter
in the stellar mass-halo mass relation, such that $\sim10\%$ of
their satellite systems have only three bright halos with
$\vmax>30$~km/s.  Likewise, we find that $\sim10\%$ of the ELVIS
realizations contain a MW without any massive failures, assuming
a scatter of 2~dex.

We caution that our analysis has ignored errors in the
stellar masses of the dwarf galaxies.  If these errors are
systematically correlated such that the majority of the stellar
masses are underestimated (overestimated), our best-fit
$\alpha$ overestimates (underestimates) the true
log-slope at a given scatter.  Furthermore, while \citetalias{Behroozi2013}
demonstrated that a single $\mstar-\mpeak$ relation describes
centrals and satellites well, it is in principle possible that
the relationship varies with environment.  If so, simulations
of isolated dwarfs (i.e. those far from any MW-size hosts) should
not necessarily match the best-fit $\alpha(\sigma$) derived here.
However, assuming an $\mstar(\mpeak) \propto \mstar^\alpha$
that is independent of environment and that behaves like a pure
power-law in the stellar mass regime explored here, such simulations
should approximately match the results presented here in order to
reproduce galaxy counts in the LG.  Finally, we explicitly note that our
models are \emph{not} expected to hold for $\mstar\gtrsim10^9\msun$,
where large-volume surveys have constrained the scatter to be
$\sigma \lesssim 0.2$~dex.

\section{Conclusions}
\label{sec:conclusions}
In this work, we have explored two-parameter models for assigning stars to
low-mass ($\mhalo \lesssim 10^{11}\,\msun$) dark matter halos in the collisionless
ELVIS simulations.  We fix the $\mstar-\mhalo$ relation to that of
\citet{Behroozi2013} for $\mhalo\ga 10^{11}\msun$ and allow the scatter $\sigma$
in this relation to vary between 0 and 2 dex, then constrain the best-fit
log-slope $\alpha$. Briefly, we find that:

\begin{itemize}
\item As the assumed scatter increases, the median relation must
become increasingly steep to avoid overproducing the observed
stellar mass functions of the MW, M31, and the nearby Local Field.

\item The best-fit log-slope $\alpha$ is significantly steeper than
the $z = 0$ slope predicted by \citetalias{Behroozi2013}, which
overproduces local galaxy counts \citep[see][]{ELVIS}.  In contrast,
the best-fit zero-scatter relation nearly matches that predicted in
\citet{ELVIS}, who derived $\alpha$ from the faint-end slope of
the SMF measured by the GAMA survey \citep{Baldry2012}.  Finally,
we find slightly shallower log-slopes than those quoted by
\citet{Moster2013}, particularly at $z = 0$, even when compared to
the field.  However, their log-slope at $z \sim 1$, roughly the median
infall redshift of subhalos around MW-size hosts \citep{Wetzel2015}, is
only slightly steeper than our measured best-fit $\alpha_{\rm sats}$.

\item At the best-fit $\alpha$, the average number of massive failures,
both within $300$~kpc of the MW and in the LF, decreases with increasing
$\sigma$.  After mocking the effects of baryonic mass loss, we find that
the MW satellite system may be only a $\sim15\%$ outlier, if there are
$\sim2$~decades of scatter about the median $\mstar-\mhalo$ relation.

\item Mass-dependent scatter produces qualitatively similar results to
mass-independent models but leads to even steeper relations for an
equivalent scatter at $\mpeak = 10^{10}\msun$, as smaller galaxies
are more likely to appear in the complete regime for growing scatter
models.

\item Currently, simulations are able to reproduce the upward scatter that
solves TBTF (by placing bright galaxies in small halos), but overall, are
thus far unable to reproduce the downward scatter (placing faint galaxies
in overly-massive halos) necessary to avoid overproducing the observed
stellar mass functions.  However, this ignores changes to the internal
structure of (sub)halos and furthermore assumes that $\mstar(\mhalo)$ can
be modeled as scatter about a median power-law.
\end{itemize}

Our results provide a method for simulators to check whether their
resultant $\mstar-\mhalo$ relation can match counts in the LG.  Specifically,
simulators should 1) fit a power-law to $\mstar-\mpeak$, normalized to the
relation in Equation~\ref{eqn:behroozi1}, to find $\alpha$; 2) measure the
scatter about that relation to determine $\sigma$ and; 3) compare with the
constraints in Figure~\ref{fig:kappagrids}.  If $\mstar-\mpeak$ is well-fit
by a power-law, but the resultant $(\alpha,\sigma)$ lies well outside the
contours provided here, then the simulations will not reproduce galaxy counts
in the LG.  Simulations that are not well fit by a power-law in $\mpeak$
are not constrained by our method.

Our results are largely independent of the exact model used
to set $\mstar$, and hold for the constant and growing scatter
models.  Together, they suggest that scatter in $\mstar-\mhalo$
\emph{alone} is unlikely to explain the TBTF problem, but that it may
contribute to the eventual solution, along with ram-pressure stripping
and internal feedback processes.  Nonetheless, it is encouraging for
the $\Lambda$CDM paradigm of galaxy formation that scatter in
$\mstar-\mhalo$, which simulations predict to at least some degree,
helps to eliminate the massive failures.  As yet, however, there
remains no direct evidence for the required scatter.

\vskip1cm

\noindent {\bf{Acknowledgments}} \\
The authors thank Michael Cooper, Marla Geha, Coral Wheeler,
Jose O\~{n}orbe, Ferah Munshi, Erik Tollerud, Frank van den Bosch,
Andrew Wetzel, Massimo Ricotti, Phil Hopkins, Justin Read, and the
anonymous referee for valuable comments that have improved the manuscript.
We also thank Aaron Dutton for providing the NIHAO data.

Support for SGK was provided by NASA through Einstein Postdoctoral Fellowship
grant number PF5-160136 awarded by the Chandra X-ray Center, which is operated
by the Smithsonian Astrophysical Observatory for NASA under contract NAS8-03060.
MBK acknowledges support from the National Science Foundation (grant AST-1517226)
and from NASA through HST theory grants (programs AR-12836 and AR-13888) awarded
by the Space Telescope Science Institute (STScI), which is operated by the
Association of Universities for Research in Astronomy (AURA), Inc., under NASA
contract NAS5-26555.

We also acknowledge the support of the \emph{Greenplanet}
cluster at UCI, where much of the analysis was performed.  This work also made use
of \texttt{matplotlib} \citep{Hunter:2007}, \texttt{Astropy} \citep{Astropy},
\texttt{numpy} \citep{numpy}, \texttt{scipy} \citep{scipy}, and \texttt{ipython}
\citep{ipython}.  This research has made use of NASA's Astrophysics Data System.

\bibliographystyle{mnras}
\bibliography{OrganizedChaos}

\label{lastpage}
\end{document}